\documentclass[10pt,journal]{IEEEtran}
\ifCLASSINFOpdf
\else
\fi
\usepackage{comment}
\usepackage{adjustbox}
\usepackage{graphicx}
\graphicspath{{./figs/}}
\usepackage{balance}
\usepackage{xspace}
\usepackage{booktabs}

\usepackage{enumitem}
\usepackage{multirow}
\usepackage{amsmath}
\usepackage{amssymb}
\usepackage{mathtools}
\usepackage{pifont}

\usepackage{tikz}
\usepackage{wasysym}
\usepackage{textcomp}

\newcommand{\Rmnum}[1]{\expandafter\@slowromancap\romannumeral #1@}

\usepackage{algorithmic}
\usepackage[linesnumbered,ruled]{algorithm2e}

\usepackage{hyperref}

\DeclareMathOperator*{\argmin}{arg\,min}

\usepackage[sort, numbers]{natbib}

\newcommand{\SI}[1]{#1}
\newcommand{\us}{}

\newcommand{\eg}{\hbox{{e.g.}}\xspace}
\newcommand{\ie}{\hbox{{i.e.}}\xspace}

\usepackage{caption}
\usepackage{subcaption}
\usepackage{comment}
\usepackage{amsmath}
\DeclareMathOperator*{\argmax}{arg\,max}

\usepackage{booktabs}
\usepackage{xcolor}

\begin{document}

\title{With Great Dispersion Comes Greater Resilience: Efficient Poisoning Attacks and Defenses for Linear Regression Models}

\author{\IEEEauthorblockN{Jialin Wen\IEEEauthorrefmark{1},
		Benjamin Zi Hao Zhao\IEEEauthorrefmark{2}, 
		Minhui Xue\IEEEauthorrefmark{3},
		Alina Oprea\IEEEauthorrefmark{4},
		Haifeng Qian\IEEEauthorrefmark{1}
	} \\
	\IEEEauthorblockA{\IEEEauthorrefmark{1}East China Normal University, China\\
		\IEEEauthorrefmark{2}The University of New South Wales and CSIRO-Data61, Australia\\
		\IEEEauthorrefmark{3}The University of Adelaide, Australia\\
		\IEEEauthorrefmark{4}Northeastern University, USA
	}
	\vspace{-10mm}
	\thanks{Haifeng Qian (hfqian@cs.ecnu.edu.cn) and Minhui Xue (jason.xue@adelaide.edu.au) are the corresponding authors of this paper.}
}

\maketitle

\begin{abstract}
With the rise of third parties in the machine learning pipeline, the service provider in ``Machine Learning as a Service'' (MLaaS), or external data contributors in online learning, or the retraining of existing models, the need to ensure the security of the resulting machine learning models has become an increasingly important topic. The security community has demonstrated that without transparency of the data and the resulting model, there exist many potential security risks, with new risks constantly being discovered. 

In this paper, we focus on one of these security risks -- \textit{poisoning attacks}. Specifically, we analyze how attackers may interfere with the results of regression learning by poisoning the training datasets. To this end, we analyze and develop a new poisoning attack algorithm. Our attack, termed \textit{Nopt}, in contrast with previous poisoning attack algorithms, can produce larger errors with the same proportion of poisoning data-points. 
Furthermore, we also significantly improve the state-of-the-art  defense algorithm, termed TRIM, proposed by Jagielsk et al.~(IEEE S\&P 2018), by incorporating the concept of probability estimation of clean data-points into the algorithm. Our new defense algorithm, termed \textit{Proda}, demonstrates an increased effectiveness in reducing errors arising from the poisoning dataset through optimizing ensemble models. We highlight that the time complexity of TRIM had not been estimated; however, we deduce from their work that TRIM can take exponential time complexity in the worst-case scenario, in excess of \textit{Proda}'s logarithmic time. The performance of both our proposed attack and defense algorithms is  extensively evaluated on four real-world datasets of housing prices, loans, health care, and bike sharing services. We hope that our work will inspire future research to develop more robust learning algorithms immune to poisoning attacks. 
\end{abstract}


\section{Introduction} 
\label{sec:introduction}

\begin{figure}[t]
\centering
\includegraphics[height=4cm,width=\linewidth]{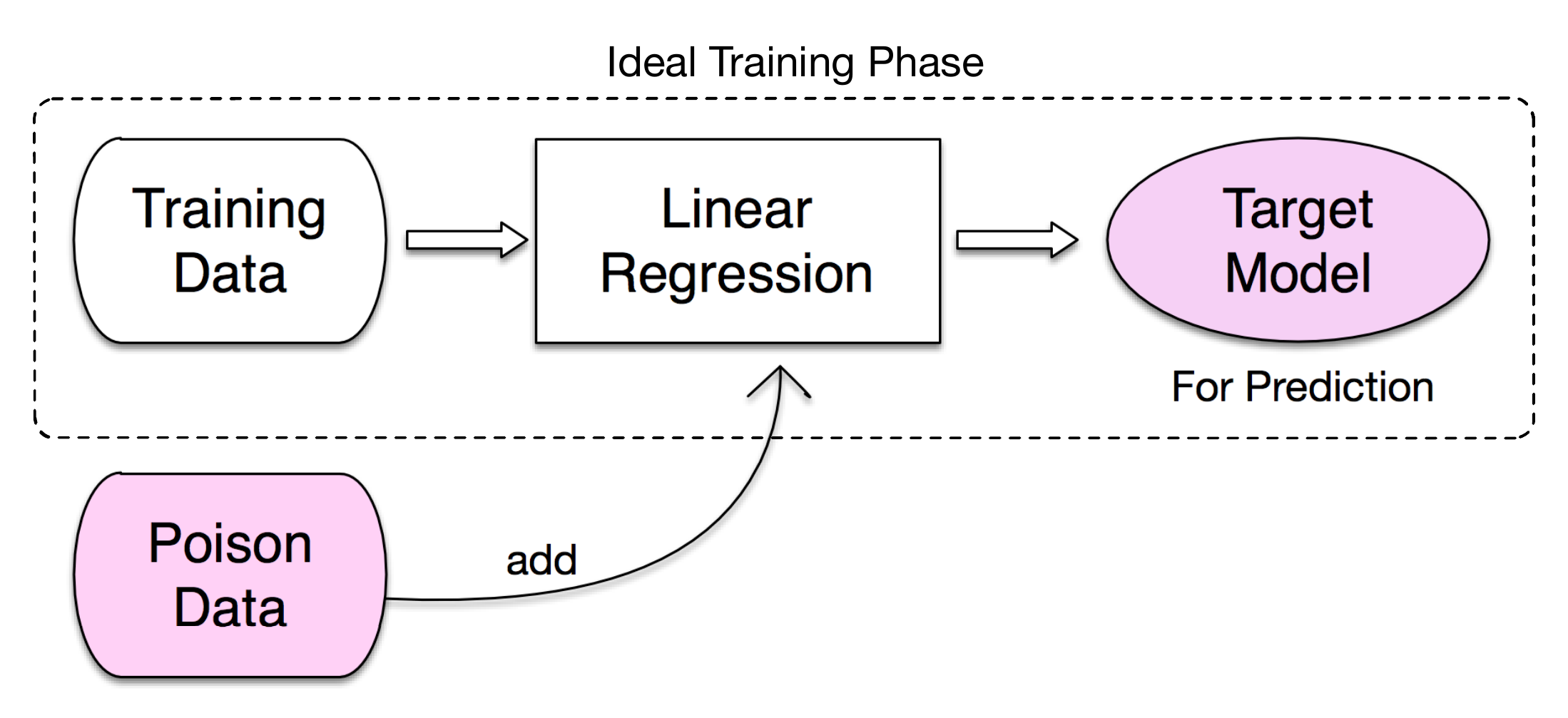}
\caption{The poisoning attack on linear regression}
\label{fig:poison_concept}
\vspace{3.5mm}
\end{figure}

With the widespread adoption of Machine Learning (ML) algorithms, it has been elevated out of the exclusive use of high-tech companies~\cite{LR}. Services such as ``Machine Learning as a Service'' (MLaaS)~\cite{privacy3} can assist companies without domain expertise in ML to solve business problems with ML. However, in the MLaaS setting, there exist poisoning attacks, in which malicious MLaaS providers can either manipulate the integrity of the training data supplied by the company or compromise the integrity of the training process.
Alternatively, in a collaborative setting, whereby a model holder solicits data contributions from multiple parties for \textit{online training or retraining of an existing model}, a malicious participant may provide poisoned training samples in their submission, thereby infecting the resulting model for all parties. 
In such poisoning attacks, the attacker's objective may be to indiscriminately alter prediction results, create a denial of service, or cause specific targeted mis-predictions during test time. The attacker seeks to create these negative effects while preserving correct predictions on the remaining test samples to bypass detection. An inconspicuous attack may produce dire consequences, thus necessitating to study poisoning attacks on ML. A conceptual example of poisoning attacks is illustrated in Figure~\ref{fig:poison_concept}. 

Many poisoning attacks have been proposed and demonstrated against different ML architectures. Specially, grey-box attacks, in which the attacker has no knowledge of the training set but has an alternative dataset with the same distribution as the training set, have been proposed against Support Vector Machines (SVMs)~\cite{SVM}, Deep Neural Networks (DNNs)~\cite{DP}, Logistic Regression (LR)~\cite{logistic}, Graph-based classification~\cite{WangAttackGraphs3}, and Recommender systems~\cite{s1}.
These works have shown that poisoning attacks are effective in interfering with the accuracy of producing classifications, or recommendations by either indiscriminately altering prediction results, or causing specific mis-predictions at test time. 
However, the aforementioned attacks target models producing a label prediction; in this work, we shall focus on models that perform regression, the prediction of a numerical value. 
Thus, with a different functional objective of regression, an attacker's objective for poisoning a regression model may also differ. 
For example, an attacker may want to increase or decrease the predicted value, or may want to maximize the dispersion of the training set. Ma et al.~\cite{WB} first propose a white-box poisoning attack against linear regression, aimed at manipulating the trained model by adversarially modifying the training set. Additionally, Jagielski et al.~\cite{LR} propose several white-box and  grey-box\footnote{The assumptions of \cite{LR} is the same as ours, albeit previously mislabelled as a black-box attack.} poisoning attacks against linear regression, which aims to increase the loss function on the original training set.

\vspace{2mm}
\noindent \textbf{Our attack contributions.} 
We scrutinize poisoning attacks on linear regression by improving and redefining the attacker's objective in existing attack models and establish a new attack optimization problem for linear regression. Our new attack, termed \textit{Nopt}, is observed to be more efficient than the state of the art~\cite{LR} (IEEE S\&P 2018), termed \textit{Opt}, in maximizing the dispersion of the training set, for the same proportion of poisoned data-points. \textit{Intuitively, the key difference between Opt~\cite{LR} and Nopt is that the optimization evaluated for each subsequent poisoning point is performed on a dataset that includes all previous poisoning points, thereby creating a new poisoning point that maximizes the loss of the collective training dataset of original clean and poisoning points.}
\vspace{2mm}

To defend against poisoning attacks on regression learning, defense mechanisms have been proposed~\cite{NelsonBCJRSSTX08,RubinsteinNHJLRTT09,BiggioCFGR11,NelsonBL11,BiggioR18}. One such approach treats poisoning attack data-points as outliers, which can be counteracted with data sanitization techniques~\cite{Cretu08,HodgeA04,Paudice2018} (i.e., input validation and removal). Another approach is through robust learning~\cite{RubinsteinNHJLRTT09,LiuLVO17,LR,XuCHP17}, as learning algorithms based on robust statistics are intrinsically less sensitive to outlying training samples; the robustness can be realized through bounded losses or specific kernel functions.
Ma et al.~\cite{WB} leverage differential privacy as a defensive measure against poisoning attacks on linear regression. Jagielski et al.~\cite{LR} propose a defense algorithm against regression learning poisoning attacks, named TRIM. TRIM offers high robustness and resilience against a large number of poisoning attacks. 
\textit{We highlight that the time complexity of TRIM had not been estimated; however, we deduce from their work that TRIM can take exponential time complexity in the worst-case scenario, in excess of our defense (termed \textit{Proda})'s logarithmic time.}

\vspace{2mm}
\noindent \textbf{Our defense contributions.} We define a new defense algorithm against poisoning attacks, termed \textit{Proda}. We are the first to introduce the concept of probability estimation of unpolluted data-points into the defense algorithm. Proda demonstrates an increased effectiveness in reducing errors arising from the poisoning dataset. Additionally, the time complexity of Proda is also lower than the state-of-the-art defense of TRIM~\cite{LR} (IEEE S\&P 2018). \textit{The key insight of achieving better efficacy is that we have prior knowledge that from the sets of points we randomly sample, there must be a group of points that all belong to the unpolluted dataset. Therefore, when comparing the minimum mean squared error (MSE) values of some groups of data-points, even if there are poisoning points in the group of the smallest MSE, they must conform to the original distribution and rules of the training dataset, and will have little impact on altering the regression model.}
 \vspace{2mm}

In this paper, we systematically study the poisoning attack and its defense for linear regression models. We define a new poisoning attack on linear regression to maximize the dispersion of the training set. Additionally, we develop a new probabilistic defense algorithm against poisoning attacks, named Proda. We extensively evaluate poisoning attacks and defenses across different four regression models (Ordinary Least Squares, Ridge Regression, LASSO, Elastic-Net) trained on multiple datasets originating from different fields, including house pricing, loans, pharmaceuticals, and bike sharing services. In summary, the overall contributions of this paper are as follows. 

\begin{itemize}

\item We develop a new grey-box poisoning attack against regression models, termed \textit{Nopt}. \textit{Nopt} outperforms the state-of-the-art attack, termed Opt, proposed by Jagielsk et al.~\cite{LR} (IEEE S\&P 2018).

\item We prove that the state-of-the-art defense, termed TRIM~\cite{LR}, is estimated to have exponential time complexity in the worst-case scenario, in excess of our \textit{Proda}'s logarithmic time. 

\item We further overhaul TRIM and propose to date the most effective defense against poisoning attacks through optimizing ensemble models, termed \textit{Proda}. The performance of both our proposed attack and defense algorithms is extensively evaluated on four real-world datasets of housing prices, loans, health care, and bike sharing services.

\end{itemize}

To the best of our knowledge, we are among the \textit{first} to systematically design, develop, and evaluate the poisoning attack and defense for linear regression models. We hope that our work will inspire future research to develop more robust learning algorithms immune to poisoning attacks.
\section{Preliminaries}
\label{sec:pre}
In this section, we first introduce linear regression and then take a deep dive into defining the threat model of this paper.

\subsection{Linear Regression}
\label{sec:linear_reg_background}
Linear Regression~\cite{LR} is a supervised machine learning algorithm, frequently used to analyze complex data relationships. Linear regression uses inherent statistical features of the training dataset to quantitatively determine mutually dependent relationships between two or more variables. By learning these relationships, the resulting linear regression model can produce a numerical output on an unseen input.

Specifically, in linear regression, the model after training is a linear function $f(x,\theta)=w^T+b$, which seeks to regress the value of $y$ for a given input $x$. 
The real parameter vector $\theta=(w,b)$ of dimension $d+1$ consists of the feature weights $w$ and the bias $b$, of dimensionality $d$ and $1$, respectively. 
However, the true value is noted as $y=f(x,\theta)+e$, containing $e$, the error between the true value and the predicted value. Assuming that $e$ is Independent and Identically Distributed (IID), the mean is $0$, the variance is fixed, and that the noise term $e$ satisfies the Gaussian distribution: 
\begin{equation}
g(y_{i}|x_{i};\theta)=\frac{1}{\sqrt{2\pi }\sigma}\exp(-\frac{(y_{i}-\theta^Tx_{i})^2}{2\sigma ^2}).
\end{equation}

Then, the maximum likelihood function of the model parameters can be obtained as the product of all training sets: 
\begin{equation}
\begin{split}
L(\theta)&=\prod_{i=1}^{m}g(y_{i}|x_{i});\theta))\\
&=\prod_{i=1}^{m} \frac{1}{\sqrt{2\pi }\sigma}\exp(-\frac{(y_{i}-\theta^Tx_{i})^2}{2\sigma ^2}).
\end{split}
\end{equation}

The maximum value of $L(\theta)$ is maintained when the log function is applied, $\log L(\theta)$.

\begin{equation}
\begin{split}
\log L(\theta)&=\log\prod_{i=1}^{m} \frac{1}{\sqrt{2\pi }\sigma}\exp(-\frac{(y_{i}-\theta^Tx_{i})^2}{2\sigma ^2})\\
&=m\log\frac{1}{\sqrt{2\pi}\sigma  }-\frac{1}{\sigma^2}\cdot \frac{1}{2}\sum_{i=1}^{m}(y_{i}-\theta^Tx_{i})^2.
\end{split}
\label{eqn:log-likelihood}
\end{equation}

In order to obtain the maximum likelihood, the latter term of Equation~\eqref{eqn:log-likelihood} is to be minimized. 

We note that the maximum likelihood of linear regression can also be converted into the minimum value of the least squares, the most common mathematical form of the loss function: 
\begin{equation}
\label{eqn:linear_loss}
L(D_{tr},\theta )={\frac{1}{2}\sum_{i=1}^{m}\left ( f\left ( x_i,\theta  \right )-y_i \right )^2}+\lambda \Omega (w),
\end{equation}
where $D_{tr}$ is the training data, $\Omega (w)$ is a regularization term penalizing large weight values, and $\lambda$ is the regularization parameter used to prevent overfitting.\footnote{Equation~\eqref{eqn:linear_loss} is an instantiation of Equation~(1) in \cite{LR} for our alternative loss $E$, as we shall  explain later in Section~\ref{sec:nopt_def}.}

The primary difference between popular linear regression methods is in the choice of the regularization term. In this paper, we study the following four regression models: Ordinary Least Squares (OLS), with no regularization, 
Ridge regression, which uses $l_2-$norm regularization, LASSO, which uses $l_1-$norm regularization, and Elastic-net regression, which uses a combination of $l_1-$norm and $l_2-$norm regularization. We elaborate  on the regularization term in context of the more common minimum least squares form of the loss function.

\begin{equation}
\label{eqn:mse}
\operatorname{MSE}={\frac{1}{m}\sum_{i=1}^{m}\left ( f\left ( x_i,\theta  \right )-y_i \right )^2}. 
\end{equation}

Our proposed attack and defense hinge on its relative effectiveness in comparison with existing methods. We shall inspect effectiveness in two aspects. Firstly, the degree of poisoning by comparing the loss function of the poisoned model with the non-poisoned model when trained on the same dataset, as a successful poisoning attack, will have increased the dispersion of the points, and thus the resulting learned regression line. The specific metric used to quantify the effect of the poisoning attack will be the Mean Squared Error (MSE) (see Equation~\eqref{eqn:mse}) of the true value from the predicted value. Secondly, specific to our defense, the time complexity of the deploying defense shall be experimentally measured in seconds.

\subsection{Threat Model}
\label{sec:threat_model}

The core objective of a poisoning attack is to corrupt the learning model generated from the training phase, such that predictions on unseen data will greatly differ in the testing phase. 
However, depending on whether the goal is to produce predictions that greatly differ on specific subsets of input data, while preserving predictions on the remaining subsets, or if predictions are to be altered indiscriminately, the poisoning attack is categorized as either an Integrity attack, or an Availability attack. A similar deconstruction of attacks is found in backdoor poisoning attacks~\cite{GuBackdoor1,Chenbackdoor2,li2019invisible, saha2019hidden,salem2020dynamic}. In this work, we consider a poisoning availability attack.

\subsubsection {Attack Assumptions}
There is a sliding scale of knowledge that is assumed to be available to the attack, from white-box to grey-box and black-box attacks. 
Under the assumptions of a white-box attack, the attacker has access to the training data $D_{tr}$, the learning algorithm $L$, and the trained parameters~$\theta$. Black-box attacks have no knowledge about the internal construction of the model, with only input and output access to the model. 
However, situated between white-box and black-box, in our  grey-box setting,\footnote{There are many types of grey-box attacks; this is our own grey-box setting.} the attacker has no knowledge of the training set $D_{tr}$ but has an alternative dataset $D_{tr}^{'}$ that has the same distribution as the pristine training set $D_{tr}$. Internal to the model, the learning algorithm $L$ is known; however, the trained parameters $\theta$ are not. We do note that an attacker, can approximate $\theta^{'}$ by optimizing $L$ on $D_{tr}^{'}$~\cite{black-box,BiggioR18}. 

It is known that black-box attacks are more practical in real-world adversaries with less knowledge required about the model. However, in this work we adopt the grey-box setting. Under the grey-box setting, we assume the adversary has no information about the structure $L$ or parameters $\theta$ of linear regression, and does not have access to any large training dataset.

\subsubsection {Poisoning Rates}
\label{sec:poision_rate}
As visualized in Figure~\ref{fig:poison_concept}, a poisoning attack is performed by injecting poisoned data into the training set before the regression model is (re)trained. The influence of an attacker on the resulting model is limited by an upper bound on the proportion $\alpha = n_p/N$ of poisoned data ($D_p$ of size $n_p$) to the original clean data ($D_o$ of size $n_o$) in the training dataset ($D_N$ of size $N=n_p + n_o$)~\cite{LR}.
An attacker has complete control of the poisoning samples, as such input feature values and responses 
can be arbitrarily set within known bounds (These feature bounds may either be derived from $D_{tr}$ or $D_{tr}^{'}$, or assumed if the data is normalized.). 
Consistent with restrictions imposed by our settings of MLaaS, online learning, and retraining, prior works rarely consider poisoning rates larger than $20\%$, as the attacker is limited to being able to control only a small fraction of the training data~\cite{LR}. 
Thus, in this paper, we shall investigate poisoning rates up to a maximum of $\alpha=0.2$. 
This maximum is motivated by prior works~\cite{LR}, as poisoning rates higher than 20\% have rarely been considered, since the attacker is assumed to be capable of controlling only a small fraction of the training data. 
This is motivated by application scenarios, such as crowdsourcing and network traffic analysis, in which attackers can only reasonably control a small fraction of participants and network packets, respectively. 
Moreover, learning a sufficiently-accurate regression function in the presence of higher poisoning rates would be an ill-posed task, as the poisoning attack would be trivial~\cite{LR}. 

\subsubsection{Defense Assumptions}
We shall also be investigating defenses. To the defender, the model is a white box (as they are the model holder), the only additional item of information a defender may not know is the poisoning rate of an attacker. It is possible for the defender to derive the poisoning rate from the size of the update data provided to it. In the event of 
an inability to derive the poisoning rate, it has been argued that a poisoning rate of $\alpha=0.2$ is representative of an upper limit of poisoning attacks, and can be assumed as a worst-case scenario.

\section{Poisoning Attacks based on Optimization}\label{sec:overview}
Previous works have discussed a poisoning attack strategy, which is  applicable not only to linear regression, but also to classification algorithms. Those poisoning attacks aim to maximize the test error. However, as we have discussed, the attack objective on linear regression is different from the attack objectives for classification algorithms, with the latter seeking to only produce a specific wrong answer. 
Therefore, we define a new poisoning attack, \textit{Nopt} poisoning attack. By establishing a new poisoning optimization algorithm for linear regression, this attack will force the model to receive a more dispersed training dataset. With a more dispersed training set this will result in larger losses and/or poor convergence on the regression task, which may erode confidence in the model holder, or simply result in worse prediction confidence in practice.

\subsection{Definition of Nopt Poisoning Attack}\label{sec:nopt_def}
In this section, we define a new form of linear regression attack, the \textit{Nopt} attack. 
Previously in Section~\ref{sec:linear_reg_background}, we observed that the loss function of linear regression is the sum of squares from each point to a regression model. When more points are added, the loss  should also increase. However, when the added points are distributed in the similar manner as the pristine data, we obtain (without considering regularization):
\begin{equation}
\label{eqn:orig_dist}
\frac{L(D_N)}{L(D_o)}= \frac{N}{n_o}.
\end{equation}

That is, the ratio of the loss function $L(D_N)$ of the new training set $D_N$ to the loss function $L(D_o)$ of the original training set $D_o$ should be equal to the ratio of the size of the new training set $N$ to the size of the original training set $n_o$ when the poisoning data is distributed in a similar way as the original training set. 
Additionally, as the new training set $D_N$ is made up of the original training set $D_o$ and poisoning data set $D_p$, we formulate our attack as follows: 
\begin{equation}
\label{eqn:attack_formualtion2}
E = \left | \frac{L(D_N)}{L(D_o)}-\frac{N}{n_o} \right | =\left | \frac{L(D_o\cup D_p)}{L(D_o)}-\frac{ (n_o\cup n_p)}{n_o} \right |.
\end{equation}

As we can see from Equations~\eqref{eqn:orig_dist} and \eqref{eqn:attack_formualtion2}, when the added poisoning points are distributed in the same way as the original data, $E=0$. The further a data point is added away from the original regression line, the higher the value of $E$. Therefore, the value of $E$ is directly related to the rate of poisoning.

\subsection{Application of Nopt Poisoning Attack}\label{sec:applications}

\begin{figure}[t]
\centering
\includegraphics[height=7.2cm,width=\linewidth]{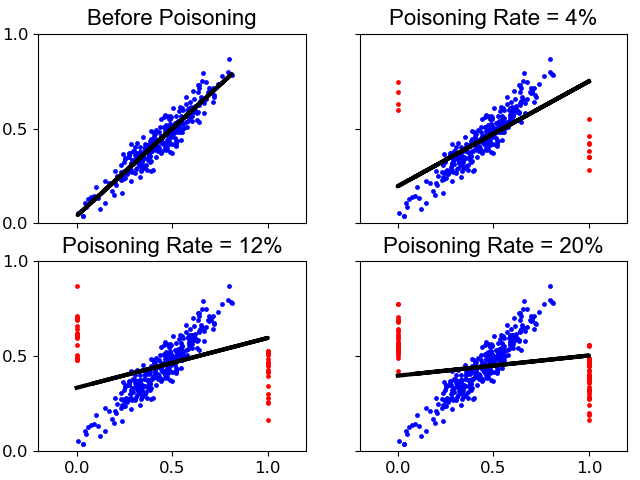}
\caption{Impact on learned regression line as a result of poisoning attacks at different poisoning rates. The original unpoisoned data is shown in blue, while poisoning data is shown in red.}
\label{visual-poisoning}
\end{figure}

Previously, Jagielski et al.~\cite{LR} established a bilevel  optimization problem to find the set of poisoning points that maximize the loss function of the original data-points. 
In this section, we compose our new poisoning attack by optimizing the objective function:
\begin{equation}
\label{eqn:optimize_attack}
\begin{aligned}
\argmax_{D_p} ~ E(D_{tr}\cup D_{p},\theta^{(p)}),\\
 \text{s.t.} \quad \theta^{(p)}\in \argmin_\theta ~ L(D_{tr}\cup D_{p},\theta  ).\\
 \end{aligned}
\end{equation}

The Nopt poisoning attack searches for poisoning data-points $D_p$ by maximizing $E$, optimizing the loss function with respect to the poisoning training dataset. 
Intuitively, the key difference between the work~\cite{LR} and Nopt is that the optimization evaluated for each subsequent poisoning point is performed on a dataset $D_{tr} \cup D_p$ that includes all previous poisoning points, thereby creating a new poisoning point that maximizes the loss of the collective training dataset of original clean and poisoning points.

\begin{algorithm}[tb]  
  \caption{Nopt poisoning attack algorithm.}  
  \label{alg:attack_framework}  
  \begin{algorithmic}[1]
    \REQUIRE  
$D=D_{tr}$ (white-box) or $D=D_{tr}^{'}$ (grey-box), $L$, $E$, the initial poisoning attack samples $D_{p}^{(0)}=(x_c,y_c)_{c=1}^{p}$, 
a small positive constant $\varepsilon$.

  \STATE $i\leftarrow 0$ (iteration counter);
 \STATE $\theta^{(i)}\leftarrow \argmin_\theta L(D\cup D_{p}^{(i)} )$;
     \REPEAT
\STATE $e^{(i)}\leftarrow E(\theta^{(i)})$;
 \STATE $\theta^{(i+1)}\leftarrow \theta^{(i)}$;

      \FOR{$c= 1, \ldots, p$}
		 \STATE $x_{c}^{(i+1)}\leftarrow $linesearch $(x_{c}^{(i)},\nabla_{x_c}E(D\cup D_{p}^{(i+1)}, \theta^{(i+1)}))$;
 		\STATE$\theta^{(i+1)}\leftarrow \argmin_\theta L(D\cup D_{p}^{(i+1)} )$;
		\STATE$e^{(i+1)}\leftarrow E(\theta^{(i+1)})$;
     \ENDFOR
     \STATE$i\leftarrow i+1$;
\UNTIL {$\left |  e^{(i)}-e^{(i+1)}\right |< \varepsilon$ }
\ENSURE  
 The final poisoning attack sample $D_{p}\leftarrow D_{p}^{(i)}$.
      \end{algorithmic}  
      \end{algorithm}

\begin{figure}[t]
\centering
\includegraphics[height=1.5cm,width=0.98\linewidth]{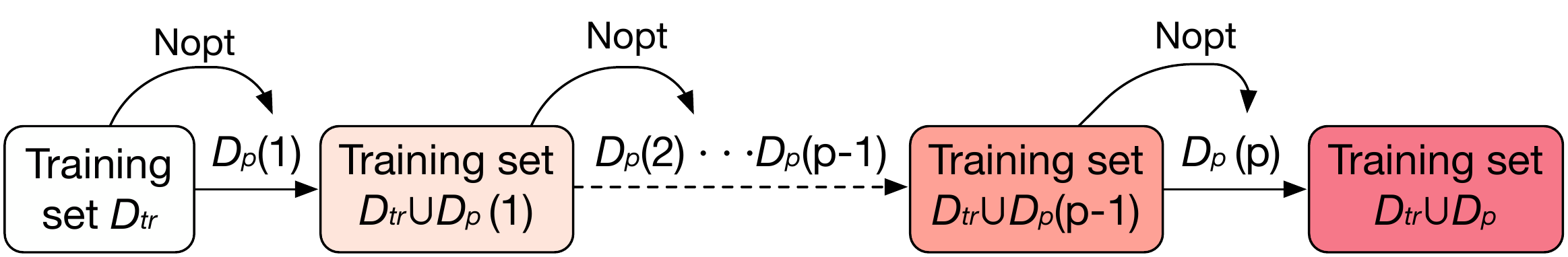}
\caption{The workflow of the poisoning attack algorithm}
\label{fig:poisoning}
\vspace{1mm}
\end{figure}

Algorithm~\ref{alg:attack_framework} outlines the Nopt poisoning attack. As our loss function takes the same form as that of \cite{LR}, the process to find the optimal points, we adopt the same gradient descent approach. In summary, vector $x_c$ is updated through a line search along the direction of the gradient from the outer objective $E$ (evaluated at the current iteration). The algorithm finishes when the outer objective $E$ yields no further changes.
Figure~\ref{fig:poisoning}, illustrates the iterative process of Nopt to find out the poisoning points.

Figure~\ref{visual-poisoning} provides a visual representation of a contrived example in which our attack is applied with different poisoning rates. 
If the abscissa of the poisoning point must be located in the feasible domain, then we can still obtain an optimal solution within the feasible domain, and $E$ will converge.

This application scenario is realistic as the attacker's goal was to create poisoning data-points to interfere with the original data-points without making the poisoning points appear abnormal, and thus compromises the secrecy of the attack. Therefore the attacker should determine the feasible domain before determining the location of the poisoning point through optimization. 
Through Equation~\eqref{eqn:attack_formualtion2}, we obtain the set of poisoning points $D_p$ with a specified poisoning degree~$E'$.

\subsection{Gradient Computation} 

The loss function we have defined in Equation~\eqref{eqn:optimize_attack} takes the same form as the loss function found in the work~\cite{LR}. Consequently, the steps of the derivation and computation of the gradients are summarized below. 

Algorithm~\ref{alg:attack_framework} reduces to a gradient-ascent task with a line search.
To compute the gradient ($\triangledown_{x_c} E(\theta^{(p)})$), whilst capturing the relationship between $\theta$ and the poisoning point $x_c$, the chain rule can be used:
\begin{equation}
\label{eqn:gradient-compute1}
\triangledown_{x_c} E=\triangledown_{x_c}\theta(x_c)^T\cdot \triangledown_{\theta} E,
\end{equation}
where the first term captures the dependency of learned $\theta$ and the point $x_c$, and the second term is the derivative of the outer objective with respect to the regression parameters $\theta$.

To solve $\triangledown_{x_c}\theta(x_c)$ in the bilevel optimization problem, the inner learning problem is replaced with its Karush-Kuhn-Tucker (KKT) equilibrium condition, namely  $\triangledown_{\theta}L(D_{tr}\cup D_{p},\theta)=0$, while searching for $x_c$. This replacement is necessary as approximations will be required to solve the inner problem, particularly when the inner problem is not convex (when the inner problem is convex, it may be solved via its closed form expression). Imposing the derivative with respect to $x_c$ satisfies this condition, $\triangledown_{x_c}{\triangledown_{\theta}L(D_{tr}\cup D_{p},\theta)=0}$. It is seen that $L$ depends explicitly on $x_c$ and implicitly through $\theta$. One final application of the chain rule produces the linear system:
\begin{equation}
\triangledown_{x_c}\triangledown_{\theta}L+\triangledown_{x_c}\theta^T\cdot\triangledown_\theta^2L=0.
\end{equation}

For our specific form of $L$ given in Equation~\eqref{eqn:linear_loss}, the derivative follows:

\begin{equation}
\label{eqn:gradient-compute2}
\begin{pmatrix}
\frac{\partial  \omega ^T}{\partial  x_c} & \frac{\partial b}{\partial x_c}
\end{pmatrix}\begin{bmatrix}
\Sigma  +\lambda g &\mu  \\ 
 \mu ^T&1 
\end{bmatrix}=-\frac{1}{n}\begin{bmatrix}
M &\omega  \\ 
\end{bmatrix},
\end{equation}
where $\Sigma =\frac{1}{n}\Sigma_ix_ix_i^T$, $\mu =\frac{1}{n}\Sigma_ix_i$, and $M=\omega x_c^T+(f(x_c)-y_c)\mathbb{I}d$. 

To jointly optimize the feature values $x_c$ associated with their responses $y_c$, we need to consider the optimization of $z_c=(x_c,y_c)$. To do this, we replace $\triangledown_{z_c}$ by $\triangledown_{x_c}$ through expanding $\triangledown_{x_c}\theta$ by incorporating derivatives with respect to $y_c$:
\begin{equation}
 \label{eqn:gradient-compute4}
 \triangledown _{z_c}\theta=\begin{bmatrix}
 \frac{\partial \omega}{\partial x_c}  &\frac{\partial \omega}{\partial y_c} \\ 
  \frac{\partial b}{\partial x_c}& \frac{\partial b}{\partial y_c}
 \end{bmatrix},
 \end{equation}
 and, accordingly, we update Equation~\eqref{eqn:gradient-compute2} as:
\begin{equation}
\label{eqn:gradient-compute3}
 \triangledown_{z_c}\theta^T=-\frac{1}{n}\begin{bmatrix}
 M &\omega  \\ 
  -x_c^T&-1 
\end{bmatrix}\begin{bmatrix}
 \Sigma  +\lambda g &\mu  \\ 
  \mu ^T&1 
\end{bmatrix}^{-1}.
 \end{equation}

Therefore, when Algorithm~\ref{alg:attack_framework} is used to implement this attack, both $x_c$ and $y_c$ are to be updated along the gradient $\triangledown_{x_c}E$ (cf. Algorithm~\ref{alg:attack_framework}, line 7).

With this, we have to use tools to perform the optimization for our Nopt poisoning attack. Following the proposal of the defense algorithm shown in Section~\ref{sec:proda_def}, in Section~\ref{sec:experiments} we will evaluate our attack in comparison to previous poisoning attacks. We shall demonstrate that our subtle change in Equation~\eqref{eqn:optimize_attack} produces larger errors compared to the previous poisoning attack on linear regression models, for the same given poisoning rates.

\section{Defense Against Poisoning Attacks}
\label{sec:proda_def}

In this section, we propose a new probabilistic defense algorithm, named \textit{Proda}, which is designed to deal with regression learning poisoning attacks. We shall analyze its time complexity and efficiency. The objective of the Proda algorithm is similar to TRIM~\cite{LR}, which is to find the original training set through a subset solving algorithm, instead of trying to identify the poisoning set of the algorithm. 
If we randomly select $ \gamma $ points, the probability of all $\gamma$ points belonging to the unpolluted training set is:
\begin{equation}
P_1=(1-\alpha)^ \gamma, 
\end{equation} 
where $\alpha$ is the poisoning rate of the training set. 
If we randomly select $\beta$ groups of $\gamma$ points, the probability that no group of points  belongs to the unpolluted training set is:
\begin{equation}
P=(1-P_1)^\beta.
\end{equation}

When $P\leq \varepsilon $, a small positive constant, 
there must be a group of points all belonging to the unpolluted training set.  We note that $ \varepsilon = 10^{-5}$ shall be used later in our experiment after considering trade-offs between accuracy and time complexity. 
This means if the value of $\beta$ follows the rules of Equation~\eqref{eqn:defense_condition}, there must be a group of $\gamma$ points that all belong to the unpolluted training set.
\begin{equation}
\label{eqn:defense_condition}
(1 - (1 - \alpha) ^ { \gamma }) ^ {\beta} \leq \varepsilon,
\end{equation}
where $\alpha$ represents the poisoning rate. So the value of $\beta$ is:
\begin{equation}
\label{eqn:beta_value}
\beta=\log_{1-(1-\alpha)^\gamma}\varepsilon.
\end{equation}

\begin{figure}[t]
\centering
\includegraphics[width=0.8\linewidth]{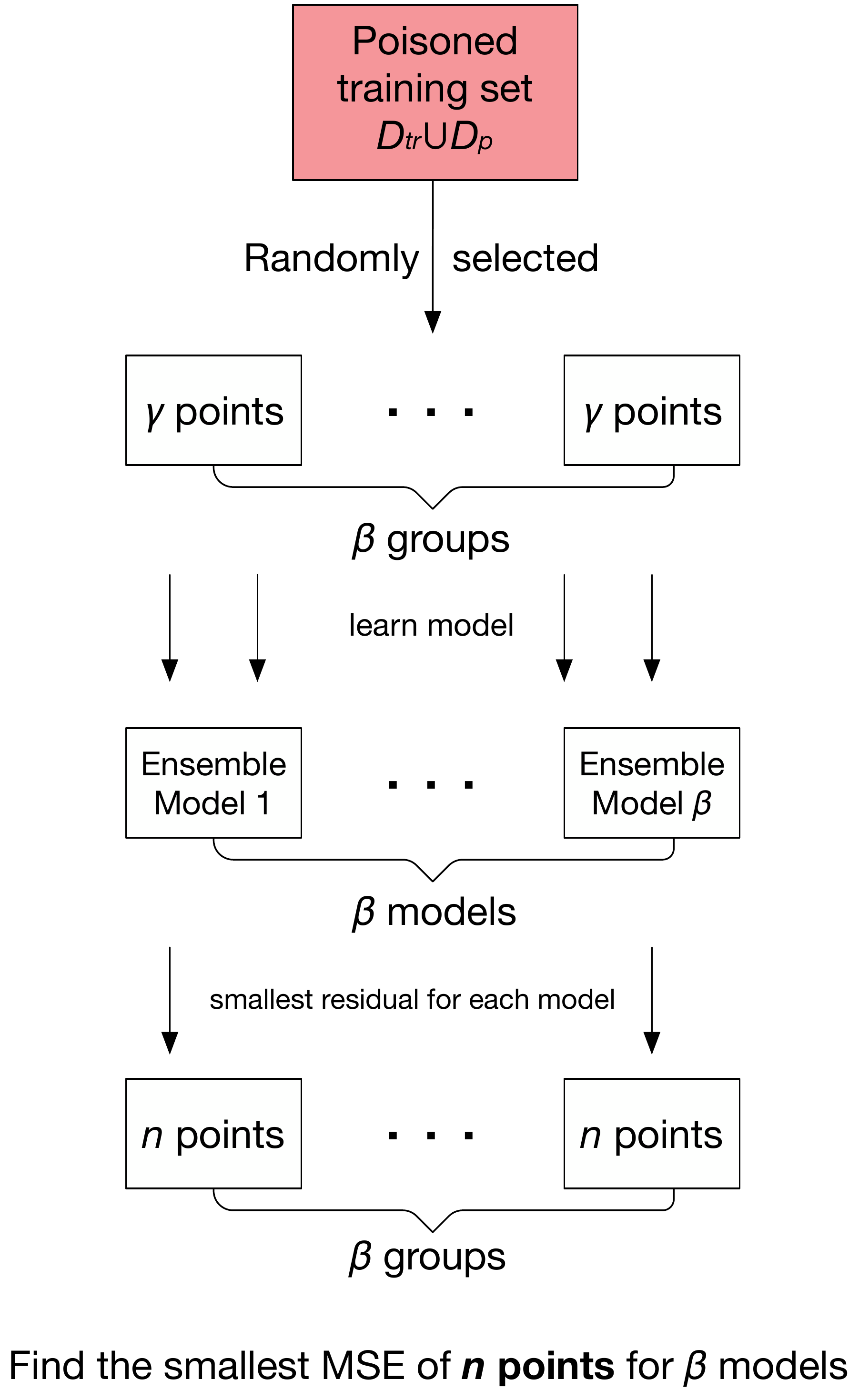}
\vspace{-1mm}
\caption{The workflow of the defense algorithm}
\label{fig:defense}
\end{figure}

The steps of Proda algorithm are illustrated in Figure~\ref{fig:defense}, and formally detailed in Algorithm~\ref{alg:defense_framework}. However, intuitively, the Proda algorithm proceeds as follows:
\begin{enumerate}
\item Given $\alpha$ and $ \gamma $, calculate the value of $\beta$.
\item Choose $\beta$ groups of $ \gamma $ points at random. (This step can ensure that there must be a group of $ \gamma $ points that all belong to the unpolluted training set).

\item Each group of $ \gamma $ points is subjected to linear regression, resulting in $\beta$ lines.
\item For each line, take the $n$ points closest to this line. 
\item Each group of $n$ points is subjected to linear regression, resulting in $\beta$ lines, and we obtain $\beta$ groups of MSEs.
\item Find $n$ points corresponding to the smallest MSE.

\end{enumerate}

\begin{figure}[tb]
\centering
\includegraphics[height=7.2cm,width=\linewidth]{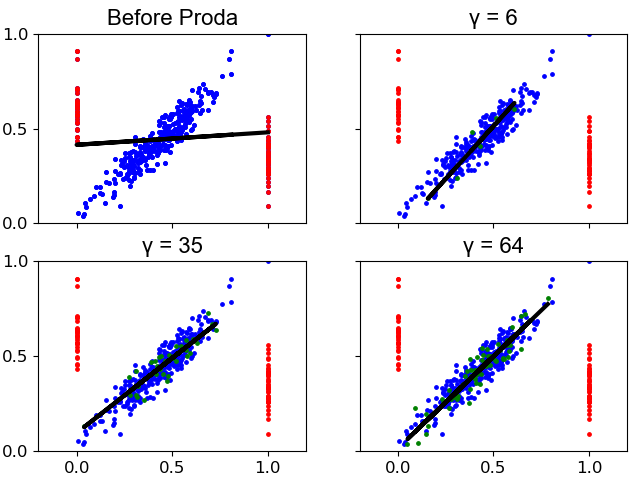}
\caption{Different parameters $\gamma$ used in the Proda algorithm. The original unpoisoned data-points are shown in blue, poisoning datapoints in red, and defensive data-points generated by the Proda algorithm in green, where green data-points are subsets of blue ones. The Proda algorithm screens out the subsets of uncontaminated datasets.} 
\label{visual-defense}
\vspace{1mm}
\end{figure}

As Proda requires the pre-selection of $\gamma$, in Figure~\ref{visual-defense}, we graphically display the effect of different selected $\gamma$ on the resulting regression line.

\begin{algorithm}[tb]  
  \caption{Proda algorithm.}  
   \label{alg:defense_framework}  
  \begin{algorithmic} [1]
    \REQUIRE
Training data $D=D_{tr}\cup D_{p}$ of $\left | D \right |=N$;  
      number of attack points $p=\alpha \cdot n$; $\gamma$; $\varepsilon$
      \STATE $\beta= \varepsilon /(1-a^\gamma$)
    \STATE $i\leftarrow$1 (iteration counter)
    \FOR {$i\leq \beta$}
    \STATE $J^{(i)}\leftarrow$ a random subset of size $ \gamma  \in \{1, \ldots, a\}$;
 \STATE $L^{(i)}\leftarrow \argmin_\theta L(J^{(i)},\theta)$;
 \STATE $list^{(i)}\leftarrow$ distance of $N$ points to $L^{(i)}$;
  \STATE $Q^{(i)}\leftarrow$ sorted$(list^{(i)})[:n]$;
  \STATE$S^{(i)}\leftarrow \argmin_\theta L(Q^{(i)},\theta )$;

      \STATE $M^{(i)}\leftarrow$ the MSE between $S^{(i)}$ and $Q^{(i)}$;
            \STATE$i\leftarrow i+1$;
\ENDFOR
     \STATE $M^{(j)}\leftarrow$min$(M^{(i)})$;
\ENSURE  
 The final optimizing sets $Q^{(j)}$.
      \end{algorithmic}  
\end{algorithm}

\subsection{Efficiency}

According to Equation~\eqref{eqn:attack_formualtion2}, we know that from the sets of points we randomly sample, there must be a group of points that all belong to the unpolluted dataset. Therefore, when comparing the MSE minimum values of $\beta$ groups, even if there are poisoning points in the group of the smallest MSE, they must conform to the original distribution and rules of the training dataset, and will have little impact on altering the regression model. Therefore, we argue that the Proda algorithm has good efficacy in mitigating the poisoning attack. An empirical evaluation on the efficiency of different defense algorithms will be demonstrated in Section~\ref{sec:experiments}.

The goal is to select $\gamma$ points that all belong to the original training set and that these points sufficiently reproduce the linear relationship of the original training set. Our algorithm can guarantee that the $\gamma$ selected points are likely to all belong to the original training set, but whether these points can represent the linear regression trend of the original training set depends on the value of $\gamma$.

It is known that for one-dimensional inputs, two points are required to define a linear relationship, between the input ($\mathbb{R}$) and the one dimensional output ($\mathbb{R}$), and for two dimensional inputs, three points, to define the linear relationship between the input ($\mathbb{R}^2$) and output ($\mathbb{R}$).
Therefore, assuming the size of uncontaminated training sets is $m=n_o$, and the feature dimensionality of the training dataset is $d$, the minimum value of $\gamma$ also needs to be at least one feature dimension greater than the input training set ($\mathbb{R}^{d}$), \ie, $\gamma \geqslant d+1$.

When $\gamma=d+1$, we can only guarantee that the line obtained by the defense algorithm is $d$-dimensional. Therefore, it is very difficult to make randomly selected $\gamma$ points in training set of $m$ points that conform to the original trends of the dataset.

Assuming that the points in the unpoisoned training set are evenly distributed, we can obtain the relationship between the MSE of the original training set $D_{tr}$ on its corresponding line $\theta_{tr}$ and the MSE of the $\gamma$ points $D_\gamma$ on its corresponding line~$\theta_{\gamma}$, as shown in Equation~\eqref{eqn:the value of gamma}:

\begin{equation}
\label{eqn:the value of gamma}
\operatorname{MSE}(D_{tr},\theta_{tr})\leq \frac{m}{\gamma}\cdot \operatorname{MSE}(D_\gamma,\theta_{\gamma}).
\end{equation}

As $\gamma$ increases, the difference between $\operatorname{MSE}(D_{tr},\theta_{tr})$ and $ \frac{m}{\gamma}\cdot \operatorname{MSE}(D_\gamma,\theta_\gamma)$ becomes smaller, defense algorithms would also be more efficient, but the corresponding time complexity will become worse (We will analyze the time complexity of the defense algorithm in detail in Section~\ref{time complexity}.). While knowledge of $\alpha$ appears to be an essential parameter for Proda, the defense algorithm will still operate for an assumed value of $\alpha$ (0.2, a safe assumption due to practical bounds~\cite{LR}). Recall that Proda creates groupings of points and selects the set of points least likely to contain poisoning points as a representative set for the dataset. Thus, if there are less poisoning points than assumed, the representative set should still only contain clean points representative of the data distribution. The largest consequence of assuming a worst case scenario for $\alpha$ is the increase in the time complexity, as we shall discuss in the next section; however, this still remains smaller than that of the competing defense of TRIM.

\subsection{Time Complexity}
\label{time complexity}
We now compute the time complexity of our proposed defense. According to Equation~\eqref{eqn:attack_formualtion2}, we know that the probability of at least one group of $\gamma$ points all belonging to the unpolluted data set is:
\begin{equation}
P_u=1-(1-(1-\alpha)^\gamma)^\beta.
\end{equation}

When $P\leq \varepsilon $, $P_u$ approaches $1$. The value of $\beta$ satisfies:
\begin{equation}
\beta \geqslant \log_{1-(1-\alpha)^\gamma}(1-P_u).
\end{equation}

That is, at least $\beta$ times of random point selection can be carried out to obtain a high probability of obtaining a group of $\gamma$ points which are all uncontaminated data.

Thus, the time complexity of Proda algorithm is:
\begin{equation}
\label{eqn:time_complexity}
T(n)= O(n)\times \log_{1-(1-\alpha)^{\gamma}}(1-P_u),
\end{equation}
where $O(n)$ is the time complexity of the linear regression algorithm. 
We note that if the feature dimensionality of the training data set is $d$, and $\gamma \geqslant d+1$, the minimum time complexity can be found as $T(n)= O(n)\times \log_{1-(1-\alpha)^{d+1}}(1-P_u)$. 

In contrast, for the time complexity of the TRIM algorithm, which we obtain by inspecting the TRIM algorithm, the worst case of TRIM involves the traversal of all $n$ training subsets; thus,  TRIM may iterate $\binom{M}{m}$ times.\footnote{No time complexity analysis was provided in TRIM's proposal~\cite{LR}.} 
Therefore, when the time complexity of TRIM is compared to our Proda algorithm, a huge improvement is expected.

\section{Experimental Analysis}
\label{sec:experiments}
Note that our poisoning attack algorithm in Section~\ref{sec:overview} 
is stated without any assumptions on the training data distribution. In practice, such information on training data is typically unavailable to attackers. Moreover, an adaptive attacker can also inject poisoning samples to modify the mean and covariance of training data. Thus, we argue that our attack algorithm results are stronger than prior works as we rely on fewer assumptions except for the work~\cite{LR} (IEEE S\&P 2018) of which the assumptions the same as ours. To evaluate the effectiveness of our attacks and defenses, we have selected two key metrics: MSE to measure the effects of poisoning as a result of the attacks and defenses (or the ``success'' of the attack), and the time complexity of each configuration; this time is computed by multiplying the number of iterations required for each configuration (we assume average of 1000 iterations per $1\mu s$ on our computer hardware).
 
Our attack and defense algorithms were implemented in Python 3.7, leveraging the NumPy and scikit-learn libraries. 
We use a conventional cross-validation method to split the datasets into three equally-sized sets for training, validation, and testing. To ensure data splitting biases are not introduced, we repeat each experiment and average results over 5 independent runs.  
 
The remainder of this section is laid out as follows. We first describe the datasets used in our experiments in Section~\ref{sec:datasets}. Followed by  Section~\ref{sec:e_attack_results}, we compare results obtained from our poisoning optimization algorithm with previous poisoning attacks on the datasets we have obtained, across four different types of regression models. Finally, we present the results of our Proda algorithm and compare it with previous defenses in Section~\ref{sec:defense_results}.

\begin{figure*}[t]
	\begin{subfigure}{0.24\textwidth}
		\centering
		\includegraphics[width=1\linewidth]{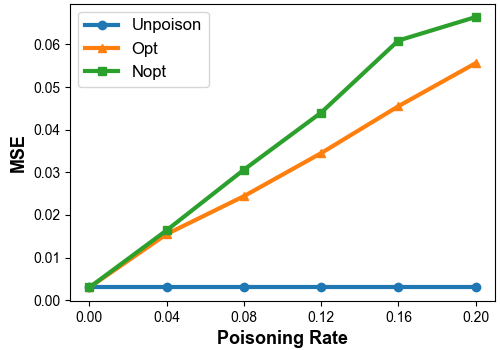}
		\caption{House Prices Dataset}
	\end{subfigure}
	\begin{subfigure}{0.24\textwidth}
		\includegraphics[width=1\linewidth]{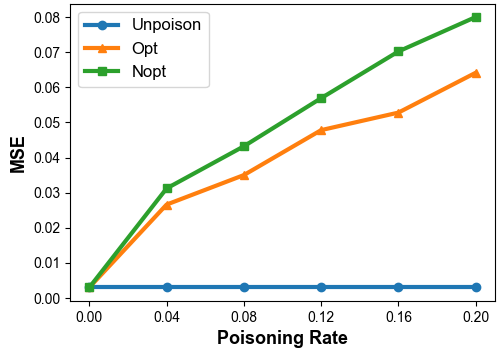}
		\caption{Loans Dataset}
	\end{subfigure}
	\begin{subfigure}{0.24\textwidth}
		\includegraphics[width=1\linewidth]{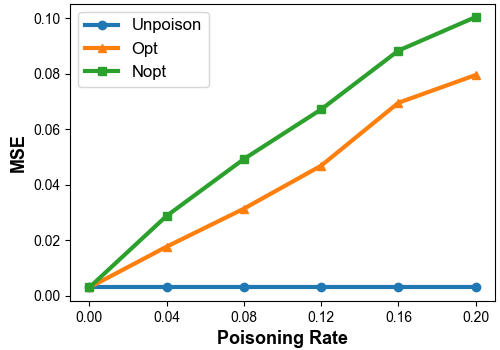}
		\caption{Pharm Dataset}
	\end{subfigure}
	\begin{subfigure}{0.24\textwidth}
		\includegraphics[width=1\linewidth]{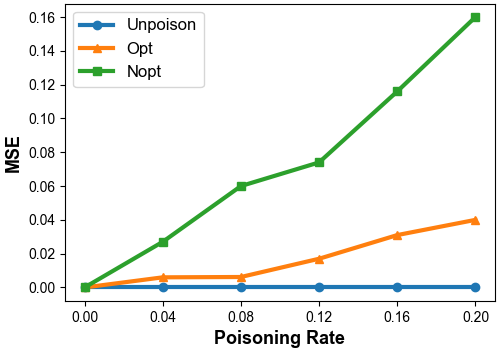}
		\caption{Bike Sharing Dataset}
	\end{subfigure}
	\vspace{-0.1cm}
	\caption{Mean Squared Error (MSE) of poisoning attacks on OLS regression on the four datasets}
    \label{fig:ols_attack}
	\vspace{-1mm}
\end{figure*}

\begin{figure*}[th]
	\begin{subfigure}{0.24\textwidth}
		\centering
		\includegraphics[width=1\linewidth]{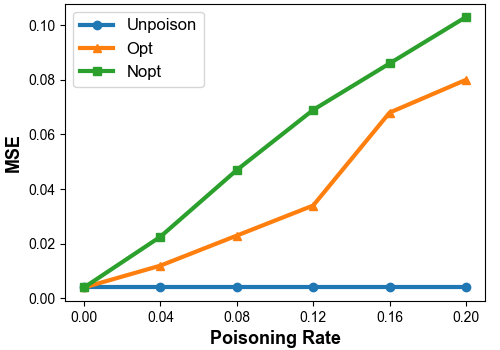}
		\caption{House Prices Dataset}
	\end{subfigure}
	\begin{subfigure}{0.24\textwidth}
		\includegraphics[width=1\linewidth]{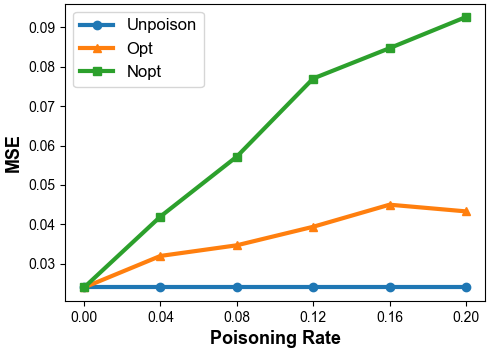}
		\caption{Loans Dataset}
	\end{subfigure}
	\begin{subfigure}{0.24\textwidth}
		\includegraphics[width=1\linewidth]{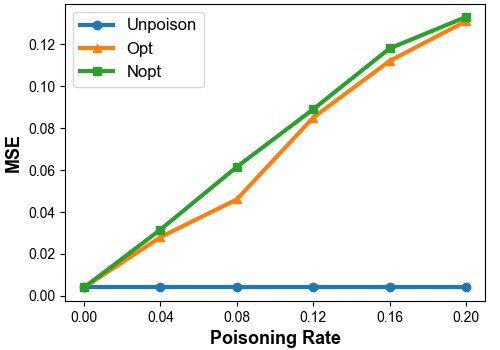}
		\caption{Pharm Dataset}
	\end{subfigure}
		\begin{subfigure}{0.24\textwidth}
		\includegraphics[width=1\linewidth]{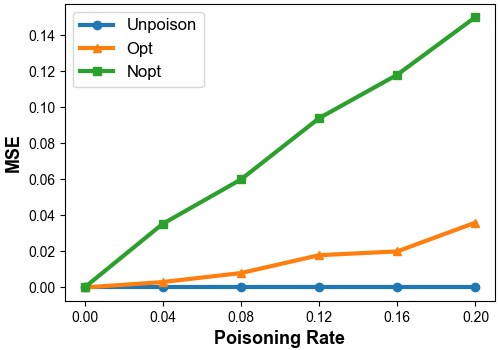}
		\caption{Bike Sharing Dataset}
	\end{subfigure}
	\vspace{-0.1cm}
	\caption{MSE of attacks on Ridge regression on the four datasets}
    \label{fig:ridge_attack}
	\vspace{-1mm}
\end{figure*}

\begin{figure*}[t]
	\begin{subfigure}{0.24\textwidth}
		\centering
		\includegraphics[width=1\linewidth]{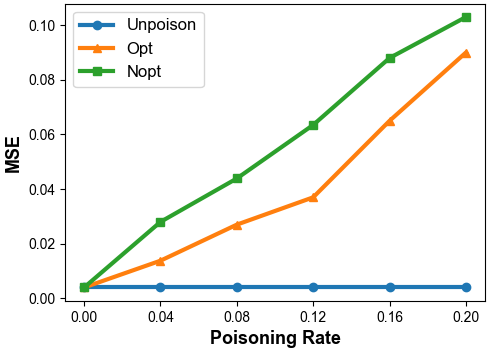}
		\caption{House Prices Dataset}
	\end{subfigure}
	\begin{subfigure}{0.24\textwidth}
		\includegraphics[width=1\linewidth]{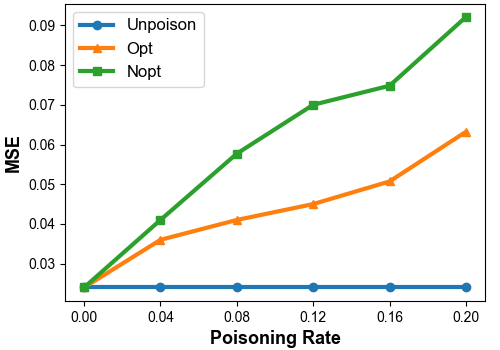}
		\caption{Loans Dataset}
	\end{subfigure}
	\begin{subfigure}{0.24\textwidth}
		\includegraphics[width=1\linewidth]{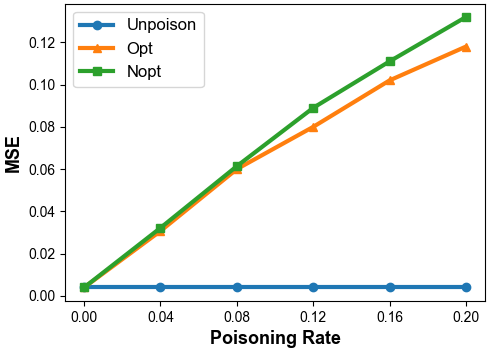}
		\caption{Pharm Dataset}
	\end{subfigure}
	\begin{subfigure}{0.24\textwidth}
		\includegraphics[width=1\linewidth]{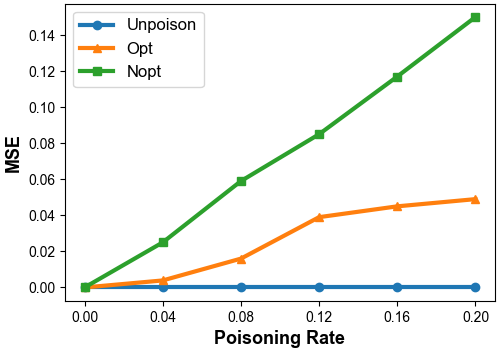}
		\caption{Bike Sharing Dataset}
	\end{subfigure}
	\vspace{-0.1cm}
	\caption{MSE of attacks on LASSO regression on the four datasets}
    \label{fig:lasso_attack}
	\vspace{-1mm}
\end{figure*}

\begin{figure*}[t]
	\begin{subfigure}{0.24\textwidth}
		\centering
		\includegraphics[width=1\linewidth]{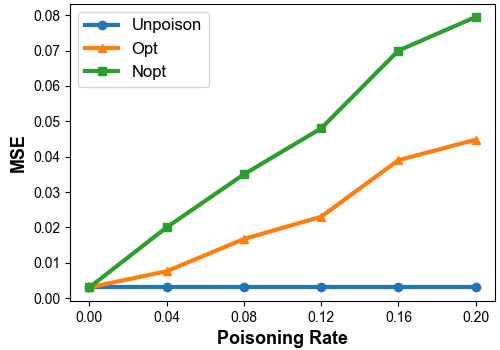}
		\caption{House Prices Dataset}
	\end{subfigure}
	\begin{subfigure}{0.24\textwidth}
		\includegraphics[width=1\linewidth]{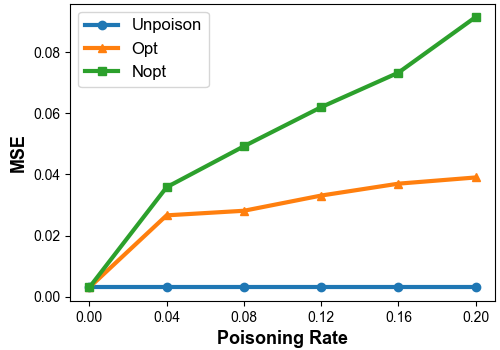}
		\caption{Loans Dataset}
	\end{subfigure}
	\begin{subfigure}{0.24\textwidth}
		\includegraphics[width=1\linewidth]{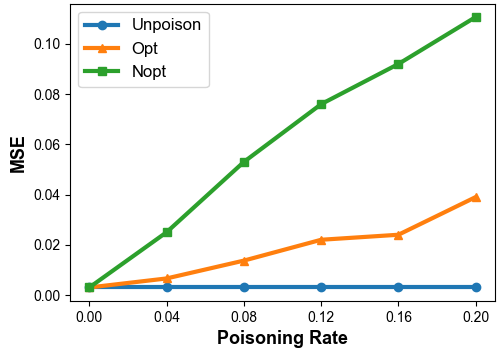}
		\caption{Pharm Dataset}
	\end{subfigure}
	\begin{subfigure}{0.24\textwidth}
		\includegraphics[width=1\linewidth]{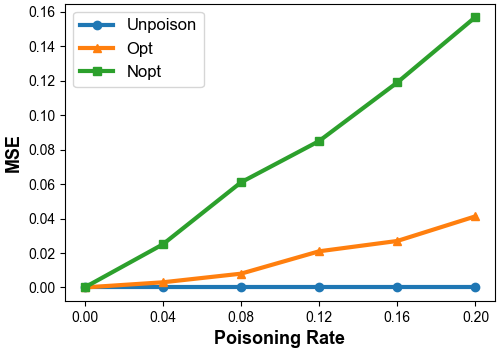}
		\caption{Bike Sharing Dataset}
	\end{subfigure}
	\vspace{-0.1cm}
	\caption{MSE of attacks on Elastic-net regression on the four datasets}
    \label{fig:enet_attack}
	\vspace{-3mm}
\end{figure*}

\subsection{Datasets}
\label{sec:datasets}

We first introduce the four publicly available datasets used in our evaluation. 

\begin{itemize}

\item \textbf{Housing Prices}~\cite{KaggleHousePrice}, a dataset used for predicting the price of the house at the time of sale given attributes of the house structure, and location information.  

\item \textbf{Loans}~\cite{KaggleLoan}, a lending dataset that seeks to estimate the appropriate interest rate of a loan given information about the total loan size, interest rate, amount of principal paid off, and the borrower's personal information such as credit status, and state of residence.

\item \textbf{Pharm}~\cite{limdi2010international}, a pharmaceuticals dataset that estimates the dosage of Warfarin for a patient depending on physical attributes of said patient, such as age, height, and weight. 

\item \textbf{Bike Sharing}~\cite{bike}, the Capital Bikeshare system dataset estimates the number of vehicles within a certain time period, given hourly information of rental bikes and environmental variables like weather. 
\end{itemize}

The aforementioned datasets, with the exception of the bike sharing dataset, have been used in previous evaluations of poisoning attacks on regression models~\cite{LR}.\footnote{We have used the datasets available at \url{https://github.com/jagielski/manip-ml}.} All datasets are pre-processed in the same manner, with categorical variables one-hot encoded, and numerical features normalized between 0 and 1. This produces 275, 89, 204, and 15 features for Housing, Loans, Pharm, and Bike Sharing, respectively. In total, each dataset contains 1460, 887383, 4683, and 17389 records, respectively. However, due to computational limitations, the only first 5, 4, 3, and 8 features of the Housing Prices, Loans, Pharm, Bike Sharing datasets, respectively, were used in the defense evaluation.

\subsection{Nopt Poisoning Attack}
\label{sec:e_attack_results}
We now perform experiments on the four selected regression datasets to evaluate our newly proposed attack. In addition, we compare our results to MSEs of the clean dataset and the optimization attack as proposed by Jagielski et al.~\cite{LR}. 
We use MSE as the metric for assessing the effectiveness of an attack, and also compute the attacks' time complexity. We vary the poisoning rate between 4\% and 20\% at intervals of 4\% with the goal of inferring the trend in attack success. Figures~\ref{fig:ols_attack},~\ref{fig:ridge_attack},~\ref{fig:lasso_attack} and~\ref{fig:enet_attack} show the MSE of each attack on OLS, Ridge, LASSO, and Elastic-net regression, respectively. We note that Jagielski et al.~\cite{LR} had only evaluated Ridge and LASSO regression. 
We plot results for the clean dataset (called Unpoison), optimization attack, proposed by Jagielski et al.~\cite{LR} (called Opt), in addition to our Nopt attack.
The horizontal coordinate is the poisoning rate $\alpha$, that is, the proportion of pollution data in the new training dataset, and the vertical coordinate is the MSE of the model trained on the poisoned dataset. Overall it can be observed from the diagrams that Nopt is able to achieve larger MSE (in comparison to Opt) for the same poisoning rate. While every configuration of the regression type, dataset and poisoning rate observes Nopt exceeding Opt, we note that Figures~\ref{fig:ridge_attack}(c) and \ref{fig:lasso_attack}(c) show a relatively similar increase of MSE between the two attacks. We also observe that OPT is more suitable for Ridge Regression and Lasso Regression on PARM Dataset; However, NOPT is seen to produce more stable attack performance.

In Table~\ref{tab:compare_attack}, we detail the specific MSEs of our new attack (Nopt) and the optimization attack proposed by Jagielski et al.~\cite{LR} (Opt). For this table, we reproduce the numerical value of the MSE at a fixed poisoning rate of $\alpha=0.2$. From Table~\ref{tab:compare_attack} we can observe that when the poisoning rate is $\alpha=0.2$, our attack (Nopt) is again consistently higher than that of the previous attack algorithm (Opt).
Our results confirm that the optimization framework we design demonstrates increased effectiveness when poisoning both different linear regression models and across datasets. The Nopt attack can achieve MSEs with a factor of $1.3$ higher than the Opt attack in the House dataset, a factor of $1.5$ higher in the Loan dataset, a factor of $1.2$ higher than Opt in the Pharm dataset, and a factor of $4.0$ higher than Opt in the bike sharing dataset. 
We note that the primary difference between NOPT and OPT is the difference in the objective function. The objective function of the OPT algorithm is determined to make the model post-poisoning deviate a maximal amount from the original model. The objective function of NOPT, however, is to perturb the original training set to become more disordered, thereby increasing the degree of dispersion of the training data set, and thus affecting the MSE of the final model on the training data set. Therefore, it is expected that the NOPT algorithm will produce more effective poisoning than the OPT algorithm, a property we have demonstrated through experimentation.

\begin{table}[t]
\centering
\caption{Comparison of MSE between Opt and our newly proposed Nopt poisoning algorithms. It can be observed that in all configurations, our attack achieves a larger MSE.} 
\label{tab:compare_attack}
\begin{adjustbox}{max width=\linewidth}
\begin{tabular}{cccc}  
\toprule
\multirow{2}{*}{Dataset}&\multirow{2}{*}{Regression}  &  \multicolumn{2}{c}{MSE after Poisoning ($\alpha=0.2$)}\cr
\cmidrule(lr){3-4} 
&& Opt & Nopt \cr
\midrule   
\multirow{4}{*}{House Prices}   &OLS  &0.055 & 0.07\\ 
&Ridge&0.07 & 0.10\\
 &LASSO&0.085 & 0.10\\
  &Elastic-net&0.04 & 0.08\\
  \midrule 
\multirow{4}{*}{Loans}   &OLS  &0.061 & 0.081\\ 
&Ridge&0.043 & 0.095\\
 &LASSO&0.062 & 0.094\\
  &Elastic-net&0.032 & 0.091\\
  \midrule 
 \multirow{4}{*}{Pharm}   &OLS  &0.077 & 0.10\\ 
&Ridge&0.134 & 0.145\\
 &LASSO&0.11 & 0.14\\
  &Elastic-net&0.031 & 0.11\\
    \midrule 
 \multirow{4}{*}{Bike Sharing}   &OLS  &0.04 & 0.16\\ 
&Ridge&0.036 & 0.15\\
 &LASSO&0.049 & 0.15\\
  &Elastic-net&0.0413 & 0.157\\
 \bottomrule  
\end{tabular}  
\end{adjustbox}
\vspace{2mm}
\end{table}

\begin{figure*}[ht!]
	\begin{subfigure}{0.24\textwidth}
		\centering
		\includegraphics[width=1\linewidth]{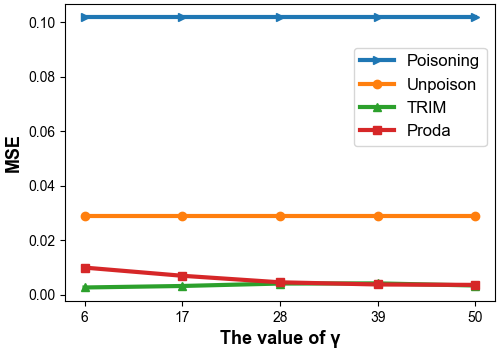}
		\caption{House Prices Dataset}
	\end{subfigure}
	\begin{subfigure}{0.24\textwidth}
		\includegraphics[width=1\linewidth]{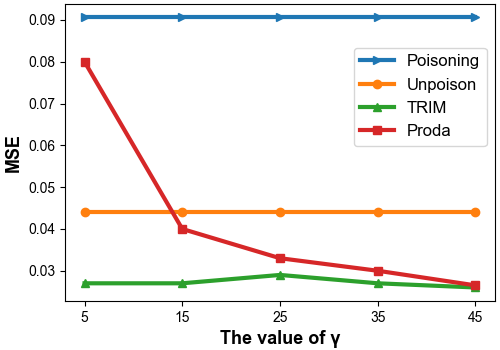}
		\caption{Loans Dataset}
	\end{subfigure}
	\begin{subfigure}{0.24\textwidth}
		\includegraphics[width=1\linewidth]{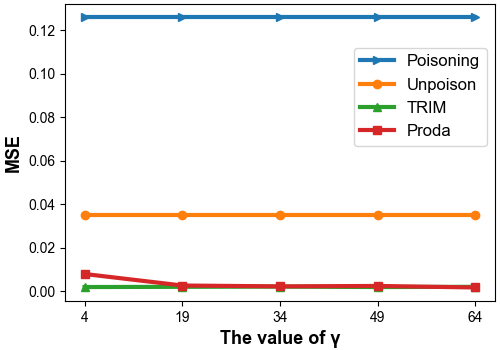}
		\caption{Pharm Dataset}
	\end{subfigure}
	\begin{subfigure}{0.24\textwidth}
		\includegraphics[width=1\linewidth]{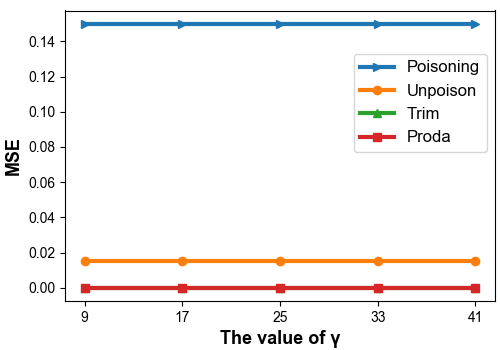}
		\caption{Bike Sharing Dataset}
	\end{subfigure}
	\vspace{-0.1cm}
	\caption{The effects of increasing $\gamma$ on the efficiency of defense algorithms on our four datasets (The poisoning algorithm is Nopt on LASSO regression, poisoning rate $\alpha=0.2$.). Both TRIM and Proda algorithms have smaller MSE results than clean data sets. When the $\gamma$ value is large, the defensive effect of the two algorithms converge. In the bike sharing dataset, two lines of TRIM and Proda overlap. The Proda algorithm  achieves a satisfying defense effect when $\gamma$ is set to be the minimum value $d+1$.  There is no general value of $\gamma$, as can be seen from Fig.~\ref{fig:gamma_defend}. For each dataset, as $\gamma$ increases, the MSE of the training set when a defense is applied will plateau. Therefore, when increasing $\gamma$ no longer significantly improving the MSE, we fix this value as the appropriate $\gamma$ value.}
    \label{fig:gamma_defend}
	\vspace{-1mm}
\end{figure*}

\begin{figure*}[ht!]
	\begin{subfigure}{0.24\textwidth}
		\centering
		\includegraphics[width=1\linewidth]{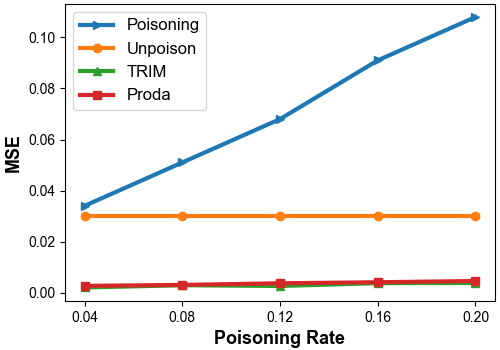}
		\caption{House Prices Dataset}
	\end{subfigure}
	\begin{subfigure}{0.24\textwidth}
		\includegraphics[width=1\linewidth]{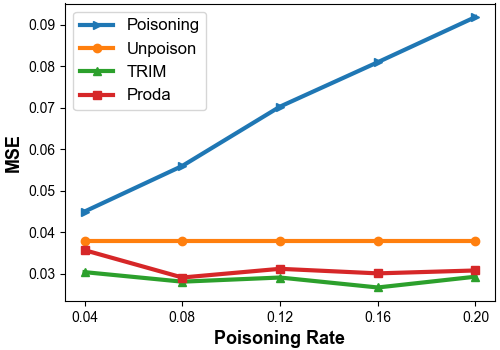}
		\caption{Loans Dataset}
	\end{subfigure}
	\begin{subfigure}{0.24\textwidth}
		\includegraphics[width=1\linewidth]{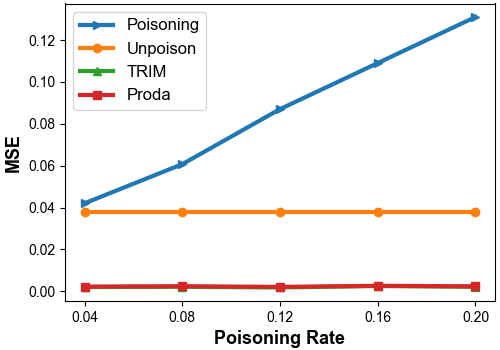}
		\caption{Pharm Dataset}
	\end{subfigure}
	\begin{subfigure}{0.24\textwidth}
		\includegraphics[width=1\linewidth]{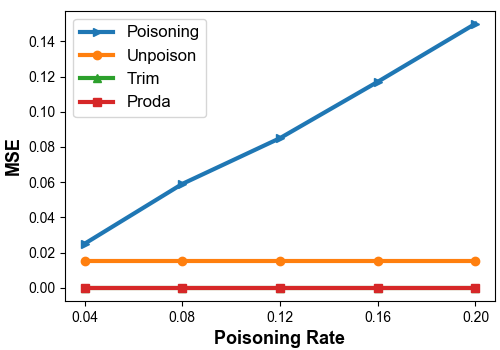}
		\caption{Bike Sharing Dataset}
	\end{subfigure}
	\vspace{-0.1cm}
	\caption{The effects of different poisoning rates $\alpha$ on the efficiency of defense algorithms on our four datasets (The poisoning algorithm is Nopt of LASSO regression, $\gamma=50$.) Both Proda and TRIM algorithms display good defensive performances at various poisoning rates. In the house price dataset, pharm and bike dataset, the two trends overlap. 
	}
	\label{fig:rate_defend}
		\vspace{-3mm}
\end{figure*}

\subsection{Proda Defense Algorithm}
\label{sec:defense_results}

In this section, we evaluate if our Proda algorithm can effectively defend against the optimized attack (Nopt), the attack that produced the highest MSE in Section~\ref{sec:e_attack_results}.

We shall use two measures in the experiment: the difference of MSE and time complexity. We evaluated the MSE among the resulting dataset of the Proda algorithm, the resulting dataset of TRIM algorithm, and the clean dataset. We also evaluated the time complexity of our defense by measuring the running time of the algorithm. For the probability algorithm, different $\gamma$ values will produce different operation results and time complexities; thus our experiments may also evaluate results for different values of~$\gamma$, however unless otherwise stated, the default value of $\varepsilon$ in our experiments is $10^{-5}$. 

\subsubsection{Defense Efficiency}
When the experiment was performed, $\gamma$ and $\alpha$ are assumed known to be the defender (Recall from Section~\ref{sec:poision_rate}, $\alpha$ is known by the defender, as they can assume that every sample submitted to the learning process is likely to be malicious; in the event that the $\alpha$ is unknown and cannot be found by the defender, the defender can assume an upper bound of 0.2 poisoning rate.). 
Therefore in specific cases, $\beta$, the intermediate variable, would not change. So, if $\gamma$ is given, MSE varies only with the poisoning rate (Figure~\ref{fig:rate_defend}, for a fixed $\gamma=50$). If a poisoning rate $\alpha$ is given, MSE varies only with $\gamma$ (Figure~\ref{fig:gamma_defend}, for a fixed $\alpha=0.2$).
\par

We remark that the value of $\gamma$ has been set to a specific value instead of a percentage of the training set. Empirically, we observed the selection of $\gamma$ had no clear correlation with the size of the training sets, instead the value of $\gamma$ is more strongly associated with the number of training set features. Specifically, we observed that the minimum value of $\gamma$ should be greater than the number of training set features ($d+1$). Consequently, we have analyzed in more detail the relationship between $\gamma$ and the number of features in the training set.  
Figure~\ref{fig:gamma_defend} shows the resulting MSE of the poisoning attack when setting different values of $\gamma$ for the defense algorithm. 
We can see that when $\gamma$ is small, the defense efficacy of Proda is less than that of TRIM (with the exception of the bike sharing dataset, where the MSE is the same), but with an increase of $\gamma$, the effectiveness of Proda in removing the influence of poisoning samples increases, however at the cost of additional computational time (see Table~\ref{tab:attack_compare}). 
As we have previously described in Section~\ref{sec:proda_def}, Proda seeks to find a subset of data that contains only clean samples, as the presence of poisoning points will increase the MSE of the resulting regression (in the overall model and in each subset evaluated by Proda), poisoning points that greatly increase the dispersion of the subset (and thus MSE) will be discarded by Proda. 
In Figure~\ref{fig:gamma_defend}(b), when $\gamma$ is equal to twice the number of features, the MSE of the probability algorithm result is lower than the MSE of the clean dataset, while the remaining datasets observe lower MSE compared to the clean dataset for all values of~$\gamma$. This indicates that a proper $\gamma$ can induce not only an effective defense result observed, but it also has assisted in the generalization of the model. We assert that the result is obtained as Proda also removes poor training points that may exist in the clean dataset. Interestingly, Proda on the Loan's dataset at very small values of $\gamma$ demonstrates an MSE larger than that of the clean dataset, given that the Loans dataset contains the largest number of records. Selecting too small a set of points to act as the representative set will negatively impact the resulting regression. 
It can be seen from Figures~\ref{fig:gamma_defend} and \ref{fig:rate_defend} that the resulting MSE of TRIM algorithm is lower than the resulting MSE of the probability algorithm and lower than the MSE of the clean dataset until $\gamma$ is much larger. 
\textit{We note that we have not evaluated other defenses since Jagielski et al.~\cite{LR} have shown that TRIM, their state of the art, is capable of outperforming prior defense mechanisms.} In the bike sharing dataset, we can see from Figure~\ref{fig:gamma_defend}(d) that three lines of Unpoison, TRIM, and Proda are almost compressed to one line, indicating their performances resemble each other. Proda has achieved a satisfying  defense effect when the parameter $\gamma$ is set as a minimum threshold, \ie,  $\gamma=d+1$. Although the three lines overlap almost completely
all the way to $\gamma=41$, if we further zoom in, the performance of Proda's defense will boost as the $\gamma$ increases.

\begin{figure*}[ht!]
	\begin{subfigure}{0.24\textwidth}
		\centering
		\includegraphics[width=1\linewidth]{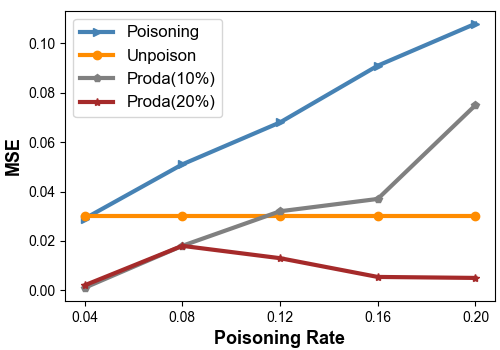}
		\caption{House Prices Dataset}
	\end{subfigure}
	\begin{subfigure}{0.24\textwidth}
		\includegraphics[width=1\linewidth]{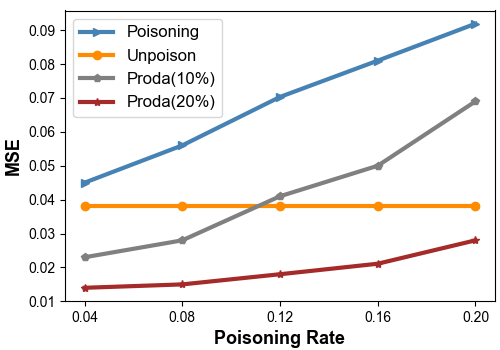}
		\caption{Loans Dataset}
	\end{subfigure}
	\begin{subfigure}{0.24\textwidth}
		\includegraphics[width=1\linewidth]{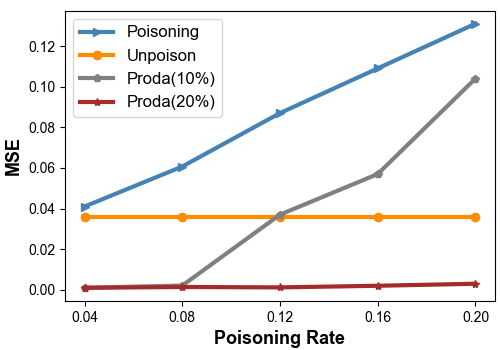}
		\caption{Pharm Dataset}
	\end{subfigure}
	\begin{subfigure}{0.24\textwidth}
		\includegraphics[width=1\linewidth]{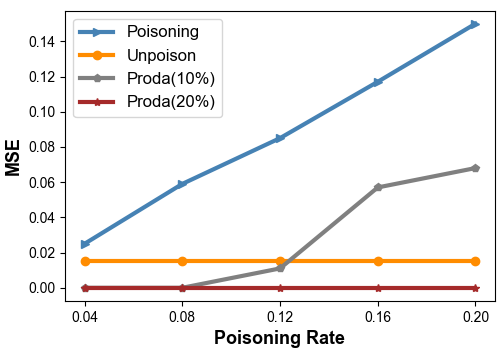}
		\caption{Bike Sharing Dataset}
	\end{subfigure}
	\vspace{-0.1cm}
\caption{Comparison of defense performance when $\alpha$ is varied for Proda on four datasets (The poisoning attack used for this evaluation is Nopt on Ridge regression with $\gamma=50$. The size of our clean training set is 300.). However, the defender does not possess knowledge about the attacker's poisoning rate, instead assuming a poisoning rate of 0.1 (10\%) and 0.2 (20\%).}
    \label{fig:alpha_defend}
	\vspace{-1mm}
\end{figure*}

\noindent \textbf{An unknown poisoning rate $\alpha$.} 
Oftentimes the defender has zero knowledge of the poisoning rate $\alpha$ used by an attacker; however, due to the construction of the Proda algorithm, it can still protect against poisoning attacks of different poisoning rates, when the defender assumes a worst case scenario of $\alpha=0.2$ (the argued largest realistic poisoning rate~\cite{LR}). In Table~\ref{tab:alpha_defense}, we demonstrate results for the Proda algorithm run with a known $\alpha$ (as previously seen in Figure~\ref{fig:rate_defend}), and the Proda algorithm when executed with an assumed $\alpha=0.2, 0.1$. From the table, we can observe that there is a small decrease in the MSE, when the assumed poisoning rate is larger than the true rate of $\alpha$. Conversely, in the event that a defender underestimates the poisoning rate $\alpha$ used by an attacker (\eg, a conservative estimate was used), we can observe from Table~\ref{tab:alpha_defense} and Figure~\ref{fig:alpha_defend} that once the real poisoning rate ($\alpha$) exceeds the defender's assumed poisoning rate, the MSE of the model exceeds that of the clean MSE. However, this increase still remains below that of the undefended MSE. This trend can be consistently observed across all datasets. Thus, it is recommended that the poisoning rate is set higher than any realistic poisoning rate for the strongest defense at the expense of time complexity (as we shall analyze in the following section); however, even a conservative estimate (for less time complexity) will still yield MSE reductions compared to no defense.

\begin{table}[t!]
\centering
\caption{Comparison of defense algorithm  when $\alpha$ is known and unknown (The poisoning algorithm is Nopt of LASSO regression, $\gamma=70$.). The effects of different poisoning rates $\alpha$ on the efficiency of defense algorithms in which the attacker does not posses knowledge of the poisoning rate $\alpha$ (The poisoning algorithm is Nopt of LASSO regression, $\gamma=70$.). Instead the defender assumes a worst case scenario of $\alpha=0.2$. Like Figure~\ref{fig:rate_defend}, both Proda and TRIM algorithms display good defensive performances at various poisoning rates. In the house price dataset and pharm dataset, the two trends overlap. In the bike sharing dataset, Unpoison, TRIM, and Proda trends overlap.} 
\label{tab:alpha_defense} 
\begin{adjustbox}{max width=\linewidth}
\begin{tabular}{cccccc}  
\toprule
\multirow{2}{*}{Dataset}&Poisoning Rate~$\alpha$&Poisoning Rate~$\alpha$ & MSE & MSE \\
&(Assumed)&(Real)&(Clean) & (Proda) & \\
  \midrule 

\multirow{9}{*}{House Prices}&\multirow{3}{*}{ 10\% }    
& 4\% & 0.03 &0.001 \\
&&12\%&0.03 &0.032 \\
&&20\%&0.03 &0.075\\
\cline{2-5}

&\multirow{3}{*}{20\%} 
& 4\% & 0.03 &0.002 \\
&&12\%&0.03 &0.013 \\
&&20\%&0.03 &0.005\\
\cline{2-5}

&4\% & 4\% & 0.03 &0.012 \\
&12\%&12\%&0.03 &0.003 \\
&20\%&20\%&0.03 &0.005\\

  \midrule 

\multirow{9}{*}{Loans}&\multirow{3}{*}{ 10\% }    
& 4\% & 0.038 &0.023 \\
&&12\%&0.038 &0.041 \\
&&20\%&0.038 &0.069\\
\cline{2-5}

&\multirow{3}{*}{20\%} 
& 4\% & 0.038 &0.014 \\
&&12\%&0.038 &0.018 \\
&&20\%&0.038 &0.028\\
\cline{2-5}

&4\% & 4\% & 0.038 &0.029 \\
&12\%&12\%&0.038 &0.029 \\
&20\%&20\%&0.038 &0.028\\

  \midrule 

\multirow{9}{*}{Pharm }&\multirow{3}{*}{  10\%  }    
& 4\% & 0.036 &0.001 \\
&&12\%&0.036 &0.037 \\
&&20\%&0.036 &0.104\\
\cline{2-5}

&\multirow{3}{*}{20\%} 
& 4\% & 0.036 &0.001 \\
&&12\%&0.036 &0.001 \\
&&20\%&0.036 &0.003\\
\cline{2-5}

&4\% & 4\% & 0.036 &0.003 \\
&12\%&12\%&0.036 &0.004 \\
&20\%&20\%&0.036 &0.003\\

  \midrule 

\multirow{9}{*}{Bike Sharing}&\multirow{3}{*}{  10\%  }    
& 4\% & 0.015 &$1.37 \times 10^{-19}$ \\
&&12\%&0.015 &0.011\\
&&20\%&0.015 &0.068\\
\cline{2-5}

&\multirow{3}{*}{20\%} 
& 4\% & 0.015 &$8.94 \times 10^{-20}$ \\
&&12\%&0.015 &$1.02 \times 10^{-19}$ \\
&&20\%&0.015 &$1.78 \times 10^{-19}$\\
\cline{2-5}

&4\% & 4\% & 0.015 &$1.96 \times 10^{-19}$ \\
&12\%&12\%&0.015 &$1.69 \times 10^{-19}$ \\
&20\%&20\%&0.015 &$1.78 \times 10^{-19}$\\

 \bottomrule 
\end{tabular}  
\end{adjustbox}
\vspace{5mm}
\end{table}

\begin{table}[t!]
\centering
\caption{Comparison of defense algorithm time performance when $\gamma$ is varied for Proda on four datasets (The poisoning attack used for this evaluation is Nopt on Ridge regression with the poisoning rate $\alpha=0.2$. The size of our clean training set is 300.).} 
\label{tab:attack_compare} 
\begin{adjustbox}{max width=\linewidth}
\begin{tabular}{ccccc}  
\toprule
\multirow{2}{*}{Dataset}&\multirow{2}{*}{Algorithm}&\multirow{2}{*}{$\gamma$} & Time Complexity & Time Complexity \\
& &&(average, $s$)  & (worst case, $s$) \\
  \midrule 
\multirow{4}{*}{House Prices}
& TRIM     & - & \SI{0.021}{\us} &$\geq5^{80}$\SI{}{\us} \\
\cline{2-5}
&\multirow{3}{*}{Proda} & 6 & \SI{0.037}{\us} & \SI{0.037}{\us} \\
&&28&\SI{5.946}{\us} &\SI{5.946}{\us} \\
&&50&\SI{806.646}{\us} & \SI{806.646}{\us}\\
 \midrule 
 \multirow{4}{*}{Loans}&
TRIM     & - & \SI{0.021}{\us} &$\geq5^{80}$\SI{}{\us} \\
\cline{2-5}
&\multirow{3}{*}{Proda} & 5 & \SI{0.028}{\us} & \SI{0.028}{\us} \\
&&25&\SI{3.041}{\us} &\SI{9.294}{\us} \\
&&45&\SI{264.318}{\us} & \SI{264.318}{\us}\\
\midrule 
 \multirow{4}{*}{Pharm}
& TRIM     & - & \SI{0.021}{\us} &$\geq5^{80}$\SI{}{\us} \\
\cline{2-5}
&\multirow{3}{*}{Proda} & 4 & \SI{0.021}{\us} & \SI{0.021}{\us} \\
&&30&\SI{9.294}{\us} &\SI{9.294}{\us} \\
&&60&\SI{7512.528}{\us} & \SI{7512.528}{\us}\\
\midrule 
 \multirow{4}{*}{Bike Sharing}
& TRIM     & - & \SI{0.021}{\us} &$\geq5^{80}$\SI{}{\us} \\
\cline{2-5}
&\multirow{3}{*}{Proda} & 9 & \SI{0.079}{\us} & \SI{0.079}{\us} \\
&&25&\SI{3.041}{\us} &\SI{3.041}{\us} \\
&&41&\SI{108.261}{\us} & \SI{108.261}{\us}\\
 \bottomrule 
\end{tabular}  
\end{adjustbox}
\vspace{5mm}
\end{table}

\begin{table}[ht]
\centering
\caption{Comparison of defense algorithm time performance when the poisoning rate $\alpha$ is varied on four datasets (The poisoning attack is Nopt on Ridge regression with a fixed $\gamma=50$. The size of our clean training set is 300.).}
\label{tab:attack_compare_alpha} 
\begin{adjustbox}{max width=\linewidth}
\begin{tabular}{cccccc}  
\toprule
\multirow{2}{*}{Dataset}&\multirow{2}{*}{Algorithm}&\multirow{2}{*}{$\alpha$} & Time Complexity & Time Complexity \\
&&&(average, $s$) & (worst case, $s$) & \\
  \midrule 

\multirow{6}{*}{House Prices}&\multirow{3}{*}{ TRIM }    & 4\% & \SI{0.021}{\us} &$\geq5^{18}$\SI{}{\us} \\
&&12\%&\SI{0.021}{\us} &$\geq5^{43}$\SI{}{\us} \\
&&20\%&\SI{0.021}{\us} &$\geq5^{80}$\SI{}{\us}\\
\cline{2-5}

&\multirow{3}{*}{Proda} & 4\% & \SI{0.143}{\us} & \SI{0.143}{\us} \\
&&12\%&\SI{12.162}{\us} &\SI{12.162}{\us} \\
&&20\%&\SI{806.646}{\us} & \SI{806.646}{\us}\\

  \midrule 

\multirow{6}{*}{Loans}&\multirow{3}{*}{ TRIM }    & 4\% & \SI{0.021}{\us} &$\geq5^{18}$\SI{}{\us} \\
&&12\%&\SI{0.021}{\us} &$\geq5^{43}$\SI{}{\us} \\
&&20\%&\SI{0.021}{\us} &$\geq5^{80}$\SI{}{\us}\\
\cline{2-5}

&\multirow{3}{*}{Proda} & 4\% & \SI{0.143}{\us} & \SI{0.143}{\us} \\
&&12\%&\SI{12.162}{\us} &\SI{12.162}{\us} \\
&&20\%&\SI{806.646}{\us} & \SI{806.646}{\us}\\

  \midrule 

\multirow{6}{*}{Pharm }&\multirow{3}{*}{  TRIM  }    & 4\%& \SI{0.021}{\us} &$\geq5^{18}$\SI{}{\us} \\
&&12\%&\SI{0.021}{\us} &$\geq5^{43}$\SI{}{\us} \\
&&20\%&\SI{0.021}{\us} &$\geq5^{80}$\SI{}{\us}\\
\cline{2-5}
&\multirow{3}{*}{Proda} & 4\% & \SI{0.143}{\us} & \SI{0.143}{\us} \\
&&12\%&\SI{12.162}{\us} &\SI{12.162}{\us} \\
&&20\%&\SI{806.646}{\us} & \SI{806.646}{\us}\\

  \midrule 

\multirow{6}{*}{Bike Sharing}&\multirow{3}{*}{  TRIM  }    & 4\% & \SI{0.021}{\us} &$\geq5^{18}$\SI{}{\us} \\
&&12\%&\SI{0.021}{\us} &$\geq5^{43}$\SI{}{\us} \\
&&20\%&\SI{0.021}{\us} &$\geq5^{80}$\SI{}{\us}\\
\cline{2-5}
&\multirow{3}{*}{Proda} & 4\% & \SI{0.143}{\us} & \SI{0.143}{\us} \\
&&12\%&\SI{12.162}{\us} &\SI{12.162}{\us} \\
&&20\%&\SI{806.646}{\us} & \SI{806.646}{\us}\\

 \bottomrule 
\end{tabular}  
\end{adjustbox}
\vspace{3mm}
\end{table}

\subsubsection{Time Complexity} 

In Section~\ref{time complexity}, we analyzed the time complexity of Proda and TRIM, and found that the worst case of Proda is superior to TRIM. In both Tables~\ref{tab:attack_compare} and~\ref{tab:attack_compare_alpha}, we detail the respective computed time complexities of Proda and TRIM, for different values of $\gamma$ (see Table~\ref{tab:attack_compare}), and different values of $\alpha$ (see Table~\ref{tab:attack_compare_alpha}). We reiterate that the time complexity indicated in these tables is computed by the number of iterations established in Section~\ref{time complexity} with an assumed average processing speed of 1000 iterations per $1\mu s$ on our computing hardware.

In Table~\ref{tab:attack_compare}, for a fixed $\alpha=0.2$, the time complexity of Proda algorithm is directly related to $\gamma$ and the number of features (recall that 5, 4, 3, and 8 features were used in House, Loan, Pharm, and Bike Sharing datasets, respectively), while the time complexity of the TRIM algorithm is related to the training data size (The size of our clean training set is 300.). 
From the average and worst-case time complexities shown in  Table~\ref{tab:attack_compare_alpha}, it can be observed that TRIM is faster than the Proda on the average case; however, our Proda algorithm provides an upper bound on the time complexity.
In Table~\ref{tab:attack_compare_alpha}, we observe time complexities for a fixed $\gamma=50$, but a varying poisoning rate~$\alpha$, we can see that the worst-case time complexity of the two algorithms increases with the increase of the poisoning rate~$\alpha$; however, as we have shown earlier in Section~\ref{time complexity}, Proda is bounded by the worst-case scenario. 

Note that TRIM algorithm uses an iterative search to find the smallest MSE subset as its defense. Suppose the size of the poisoned training set is $k$. The worst-case scenario for TRIM is to exhaust $\binom{k}{(1-\alpha)k}$ subsets to converge. In the case of a poisoning rate of $\alpha=0.2$, TRIM will not terminate until it has compared MSEs of $\binom{k}{0.8k}$ subsets ($\geqslant 1.6^{k}$ subsets, given Stirling's approximation). In contrast, Proda algorithm squarely selects the smallest MSEs of $1-\alpha k$ clean data-points from our randomly chosen subsets; hence, the time complexity of Proda algorithm is dependent only on the value of the parameter $\gamma$ we define, consequently offering a substantial reduction in the worst-case time complexity of such a defense.

\subsubsection{Effectiveness of Proda against Opt and Nopt}
Both Opt and Nopt add previously poisoned data points during the optimization phase; thus it is desirable to compare whether the two different attack algorithms will perform differently against the same defense of Proda. For all four datasets and across three values of $\alpha$ (4\%, 12\%, 20\%), it is observed in Table~\ref{tab:compare_attack1} that there is little to no difference between attack algorithms.

\begin{table}[t]
\centering
\caption{Comparison of MSE after defense between Opt and our newly proposed Nopt poisoning algorithms when the defense algorithm is the same (The  poisoning  algorithm  is  Nopt  of  LASSO regression. The size of our clean training set is 300. The defense used for this evaluation is Proda, $\gamma=70$.).} 
\label{tab:compare_attack1}
\begin{adjustbox}{max width=\linewidth}
\begin{tabular}{ccccccc}  
\toprule
\multirow{2}{*}{Dataset}&\multirow{2}{*}{$\alpha$}  &  MSE&MSE&MSE&MSE&MSE\cr
\cmidrule(lr){3-7} 
&&(Clean)& Opt & Nopt& (Opt after Proda)&(Nopt after Proda)\cr
\midrule   
\multirow{4}{*}{House Prices}   &4\%&0.03  &0.055 & 0.07&0.012&0.012\\ 
&12\%&0.03&0.07 & 0.10&0.003&0.003\\
 &20\%&0.03&0.085 & 0.10&0.005&0.005\\
  \midrule 
\multirow{4}{*}{Loans}   &4\%  &0.038&0.061 & 0.081&0.029&0.028\\ 
&12\%&0.038&0.043 & 0.095&0.029&0.029\\
 &20\%&0.038&0.062 & 0.094&0.028&0.028\\
  \midrule 
 \multirow{4}{*}{Pharm}   &4\%&0.036  &0.077 & 0.10&0.004&0.003\\ 
&12\%&0.036&0.134 & 0.145&0.004&0.004\\
 &20\%&0.036&0.11 & 0.14&0.004&0.003\\
    \midrule 
 \multirow{4}{*}{Bike Sharing}   &4\%&0.015  &0.04 & 0.16&$1.89 \times 10^{-19}$&$1.96 \times 10^{-19}$  \\ 
&12\%&0.015&0.036 & 0.15&$1.72 \times 10^{-19}$ &$1.69 \times 10^{-19}$ \\
 &20\%&0.015&0.049 & 0.15&$1.87 \times 10^{-19}$ &$1.78 \times 10^{-19}$ \\
 \bottomrule  
\end{tabular}  
\end{adjustbox}
\vspace{2mm}
\end{table}

\section{Discussion and Related Work}
In this section, we discuss the limitations of this paper and survey the related work.

\subsection{Limitations}
Supervised machine learning algorithms can solve common regression and classification problems, but the training data of supervised machine learning may potentially be manipulated by attackers seeking to interfere with the training process for their own nefarious purposes. 
For example, an attacker may add poisoning data to interfere with a supervised machine learning algorithm and likewise, a defender can use data optimization to prevent data poisoning attacks.
In data poisoning attacks, attackers may have a variety of goals to interfere the regression result. In this work we focus on maximizing the dispersion of the training set. As we have discussed earlier, an investigation into the possibility of a targeted attack is left for future work.

When defending against poisoning attacks, the Proda algorithm seeks to find a subset of points that minimize the linear regression loss function. In prior work the TRIM~\cite{LR}, a defense algorithm, has been proposed against linear regression poisoning attacks. Both TRIM~\cite{LR} and our work are based on the premise that the loss function of linear regression greatly increases with the addition of even a single poisoned point, a method that differs from other approaches which seek to classify or isolate poisoned points. What our approaches differ is that the worst-case time complexity of our defense algorithm Proda is better than TRIM~\cite{LR}. 
While we have experimentally demonstrated the effectiveness in removing the effect of poisoning attacks, it is possible that Proda can be retooled to detect the subset of poisoning points.

Our Proda algorithm has a parameter $\gamma$, which controls the precision and time complexity of the defense algorithm. The setting of the parameter $\gamma$ is data-dependent; however, we have established that $\gamma$ should not be less than one more than the dimension of the dataset $d+1$. Experimentally, we observed that a larger $\gamma$ will produce lower MSE at the expense of a higher time complexity. It can be seen from Figure~\ref{fig:gamma_defend} that when $\gamma$ approaches $d^2$, Proda algorithm can obtain basically stable defense efficiency. The specific value of $\gamma$ should be chosen based on a time complexity acceptable to the defender.

\subsection{Related Work}

\vspace{1mm}
\noindent \textbf{Poisoning attacks.} 
Data poisoning attacks are a general class of attacks that manipulate the training data of a machine-learning system such that the learned model behaves in a way dictated by the attacker. Such poisoning techniques have been studied in various applications, such as anomaly detection~\cite{RubinsteinNHJLRTT09} and email spam filtering~\cite{NelsonBCJRSSTX08}. It has also been shown that data poisoning attacks are indifferent to the underlying machine learning algorithms (SVMs~\cite{SVM}, regression~\cite{LR,ZONG,wen2020palor}, graph-based approaches~\cite{WangAttackGraphs3,44,xu2021explainability}, neural networks~\cite{GuBackdoor1,li2021hidden,9186317,li2020deep,chen2021oriole}, and federated learning~\cite{9}). While the aforementioned works primarily compromise classification tasks, there has also been work on poisoning recommender systems~\cite{10,19,39}. 
Most relevant to ours are attacks against regression tasks~\cite{logistic,LR}, whereas our attack outperforms the prior works.

\noindent\textbf{Defenses against poisoning attacks.}  
Many defense mechanisms have been proposed to defend against poisoning attacks~\cite{NelsonBCJRSSTX08,RubinsteinNHJLRTT09,BiggioCFGR11,BiggioR18}. One core approach to mitigating poisoning attacks is to recognize that the poisoned data is intentionally dissimilar to the clean training data (as to inflict the greatest change to the learning process). Due to the limitation on the proportion of poisoned data to clean data an attacker has control over, the poisoned data may be treated as outliers, and mitigated with data sanitization techniques~\cite{Cretu08,HodgeA04,Paudice2018}. The other means to defend against potentially poisoned data is to use robust learning algorithms~\cite{RubinsteinNHJLRTT09,LiuLVO17,LR,XuCHP17}. These algorithms are designed to limit the sensitivity to any single sample within the training data. Differential privacy has also been investigated as a means to mitigate data poisoning attacks~\cite{WB}. Specifically for regression models, the closest to this work is TRIM~\cite{LR}, a defense capable of managing a large number of poisoning data points. Like Proda, TRIM finds a subset of training data that minimizes the model loss, and uses this subset as a representative set to train the non-poisoned model.

Presently, the defense methods against deep learning poisoning attacks can be divided into three stages: data and feature modification, model modification, and output defense.

Data and feature modification~\cite{NelsonBCJRSSTX08,RubinsteinNHJLRTT09,BiggioCFGR11,BiggioR18} primarily refers to the processing of data or features before it is accepted as input into the model to achieve the defense objective.

A core tenet to mitigating poisoning attacks is to recognize that the poisoned data is intentionally dissimilar to the clean training data (as to inflict the greatest change to the learning process). Due to limitations to the proportion of poisoned data to clean data an attacker can control, the poisoned data is the minority class and may be treated as outliers, and mitigated with data sanitization techniques~\cite{Cretu08,HodgeA04,Paudice2018}. Another method to defend against potentially poisoned data is to use robust learning algorithms~\cite{RubinsteinNHJLRTT09,LiuLVO17,LR,XuCHP17}. These algorithms are designed to limit the model's sensitivity to any single sample within the training data. Differential privacy has also been investigated as a means to mitigate data poisoning attacks~\cite{WB}. Specifically for regression models, TRIM~\cite{LR} is a defense capable of managing a large number of poisoning data points. Like Proda, TRIM finds a subset of training data that minimizes the model loss, and uses this subset as a representative set to train the non-poisoned model.
From the perspective of preprocessing features, Shen et al.~\cite{shen} propose a method to automatically produce shielding features to identify and neutralize abnormal distributions. 

Directly protecting the model involves the modification of the deep learning model. Liu et al.~\cite{liu} propose a pruning algorithm to reduce the size of the backdoor networks by eliminating dormant neurons on pure inputs. Iandola et al.~\cite{iandola} propose a fine-tuning defense in which a potentially poisoned neural network is updated to disable backdoor triggers. DeepInspect~\cite{chen}, a detection framework, is proposed to leverage a conditional generation model to learn the probability distribution of potential triggers from the queried model to retrieve fingerprints left behind during the insertion of the backdoor. 

An output defense mitigates an attack by analyzing the output behavior of the potentially poisoned deep learning model. Yang et al.~\cite{yang} propose a loss-based method, which would trigger an accuracy check if the loss of the target model exceeded a threshold multiple times. Hitaj et al.~\cite{Hitaj} presents an integrated defense, by combining the prediction results of different models to judge the prediction categories of samples. Chandola et al's work detects poisoned inputs through Support Vector Machines (SVM) and decision trees~\cite{Chandola}. Zhao et al.~\cite{zhao3} propose the multi-task model defense and  analyze the output results of the model through data cleaning to improve the robustness of multi-task joint learning.

\section{Conclusion}\label{sec:conclusion}

We systematically study the poisoning attack and its defense for regression models in  MLaaS setting. We have proposed a modification to an attack optimization framework that requires no additional knowledge of the training process, yet produces better offensive results.
Our attack allows attackers to carry out both white-box and grey-box attacks and is capable of increasing the dispersion of the poisoned training set. However, as attackers may have  many goals in the interference of the regression outcome, the possibility of targeted attacks remains in question. 
To respond to a more powerful poisoning attack developed in this paper, we designed a probabilistic defense algorithm, Proda, which can be tuned to effectively mitigate poisoning attacks on the regression model while significantly reducing the worst-case time complexity. We highlight that the time complexity of the state-of-the-art defense, TRIM, had not been estimated; however, we deduce from their work that TRIM can take exponential time complexity in the worst-case scenario, in excess of Proda's logarithmic time. 
Finally, we hope that our work will inspire future research to develop more robust learning algorithms that are not susceptible to poisoning attacks.

\section*{Acknowledgments}
The authors affiliated with East China Normal University were, in part, supported by NSFC-ISF Joint Scientific Research Program (61961146004) and Innovation Program of Shanghai Municipal Education Commission (2021-01-07-00-08-E00101). Minhui Xue was, in part, supported by the Australian Research Council (ARC) Discovery Project (DP210102670). Alina Oprea was supported by the U.S. Army Combat Capabilities Development Command Army Research Laboratory under Cooperative Agreement Number
W911NF-13-2-0045 (ARL Cyber Security CRA). The views
and conclusions contained in this document are those of the
authors and should not be interpreted as representing the official policies, either expressed or implied, of the Combat Capabilities Development Command Army Research Laboratory
or the U.S. Government. The U.S. Government is authorized
to reproduce and distribute reprints for Government purposes
notwithstanding any copyright notation.

\bibliographystyle{IEEEtran}
{\small\bibliography{TIFS2021.bib}}

\begin{IEEEbiography}[{\includegraphics[width=1.2in,height=1.2in,clip,keepaspectratio]{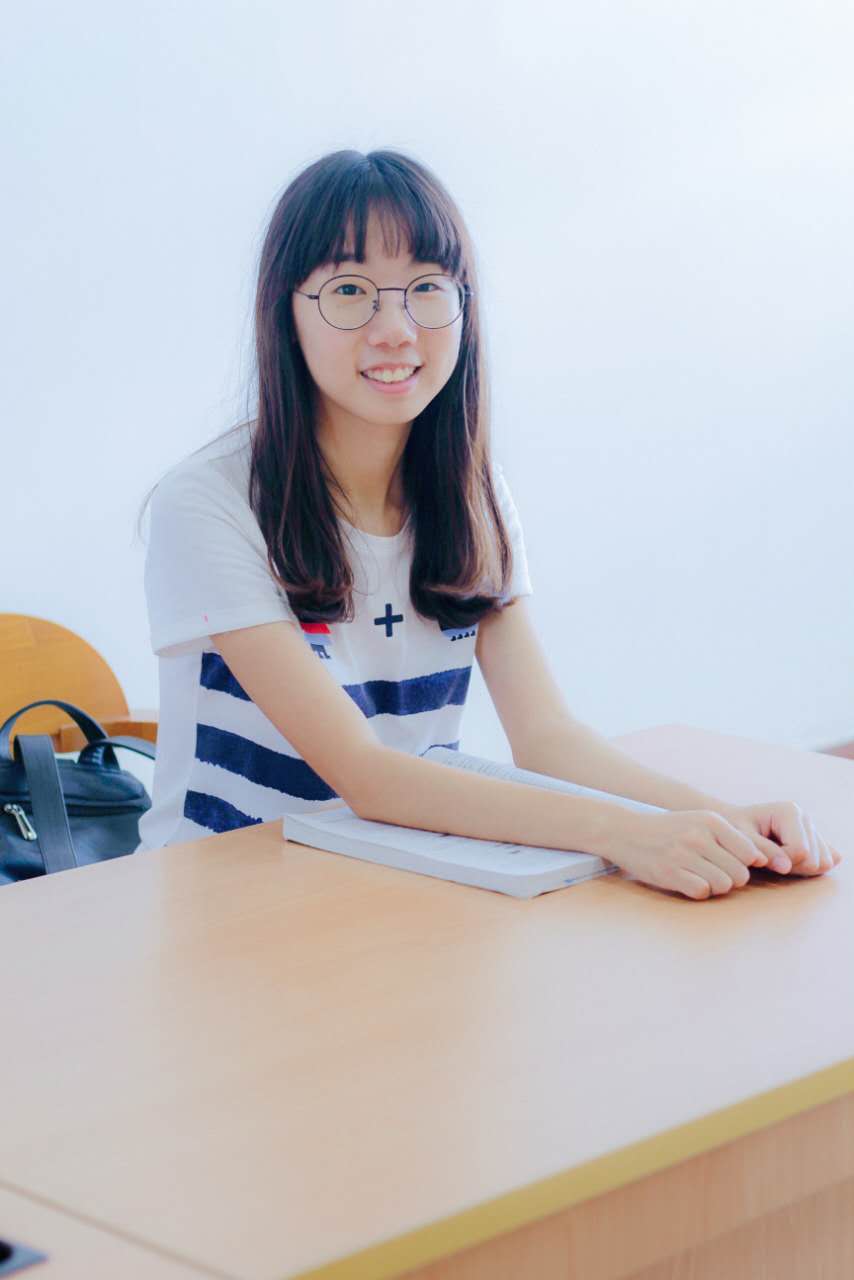}}]
{Jialin Wen} is pursuing her Master's degree at School of Computer Science of Technology of East China Normal University. She focuses primarily on the areas of machine learning and security, specifically exploring the robustness of machine learning models against various adversarial attacks.   
\end{IEEEbiography}

\begin{IEEEbiography}[{\includegraphics[width=1.2in,height=1.2in,clip,keepaspectratio]{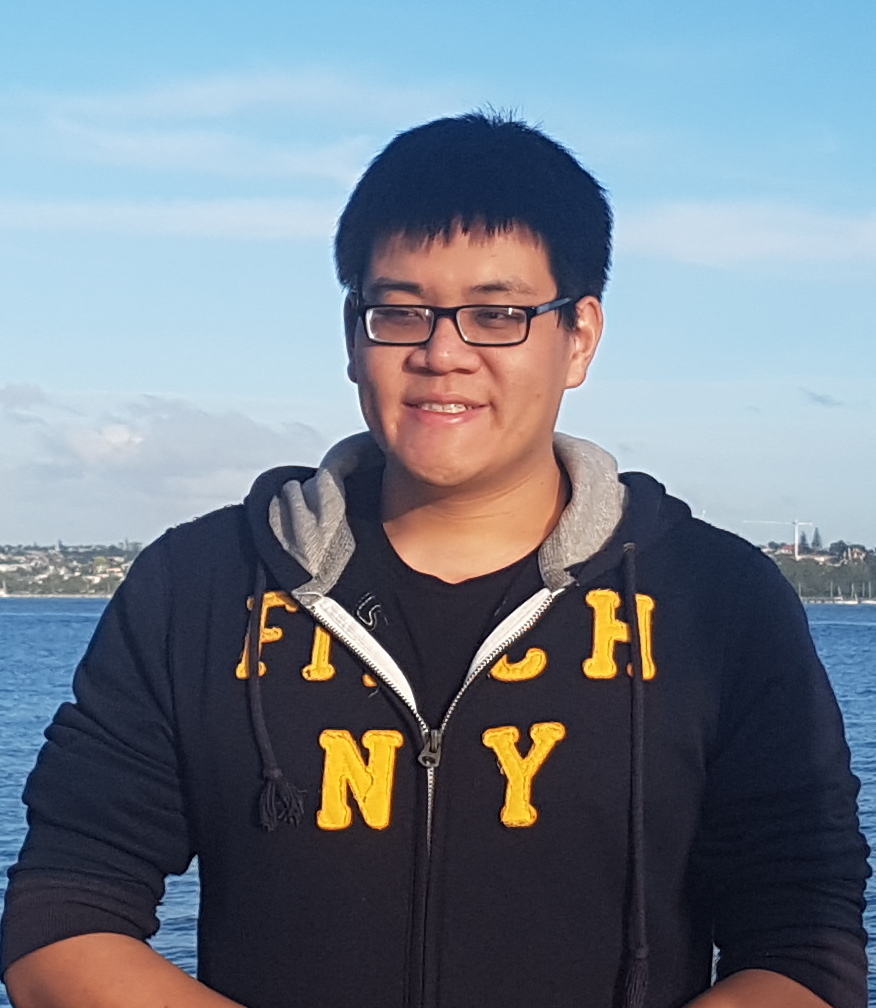}}]
{Benjamin Zi Hao Zhao} is pursuing his Ph.D. at the School of Electrical Engineering and Telecommunications at the University of New South Wales and CSIRO-Data61. His current research interests are authentication systems, and security and privacy with machine learning. His work has received the ACM AsiaCCS best paper award. 
\end{IEEEbiography}

\begin{IEEEbiography}[{\includegraphics[width=1.2in,height=1.2in,clip,keepaspectratio]{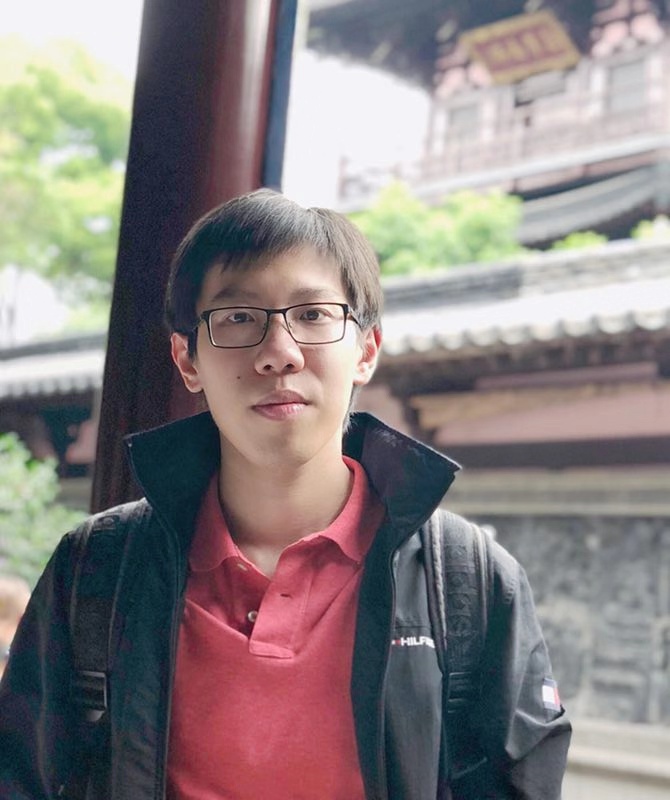}}]
{Minhui Xue} is a Lecturer (a.k.a. Assistant Professor) of School of Computer Science at the University of Adelaide. He is also an Honorary Lecturer with Macquarie University. He is the recipient of the ACM SIGSOFT distinguished paper award and IEEE best paper award, and his work has been featured in the mainstream press, including The New York Times and Science Daily. He currently serves on the Program Committee of IEEE Symposium on Security and Privacy (Oakland) 2021, ACM CCS 2021, USENIX Security 2021, NDSS 2021, ICSE 2021, ESORICS 2021, and PETS 2021 and 2020.
\end{IEEEbiography}

\begin{IEEEbiography}[{\includegraphics[width=1.05in,height=1.2in,clip,keepaspectratio]{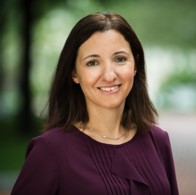}}]
{Alina Oprea} is an Associate Professor at Northeastern University in the Khoury College of Computer Sciences. She was the recipient of the Technology Review TR35 award for research in cloud security in 2011 and the recipient of the Google Security and Privacy Award 2019. She currently serves as Program Committee co-chair for the IEEE Symposium on Security and Privacy 2021. She served as Program Committee co-chair for the IEEE Symposium on Security and Privacy 2021 and 2020, as well as NDSS 2019 and 2018.
\end{IEEEbiography}

\begin{IEEEbiography}[{\includegraphics[width=1.2in,height=1.2in,clip,keepaspectratio]{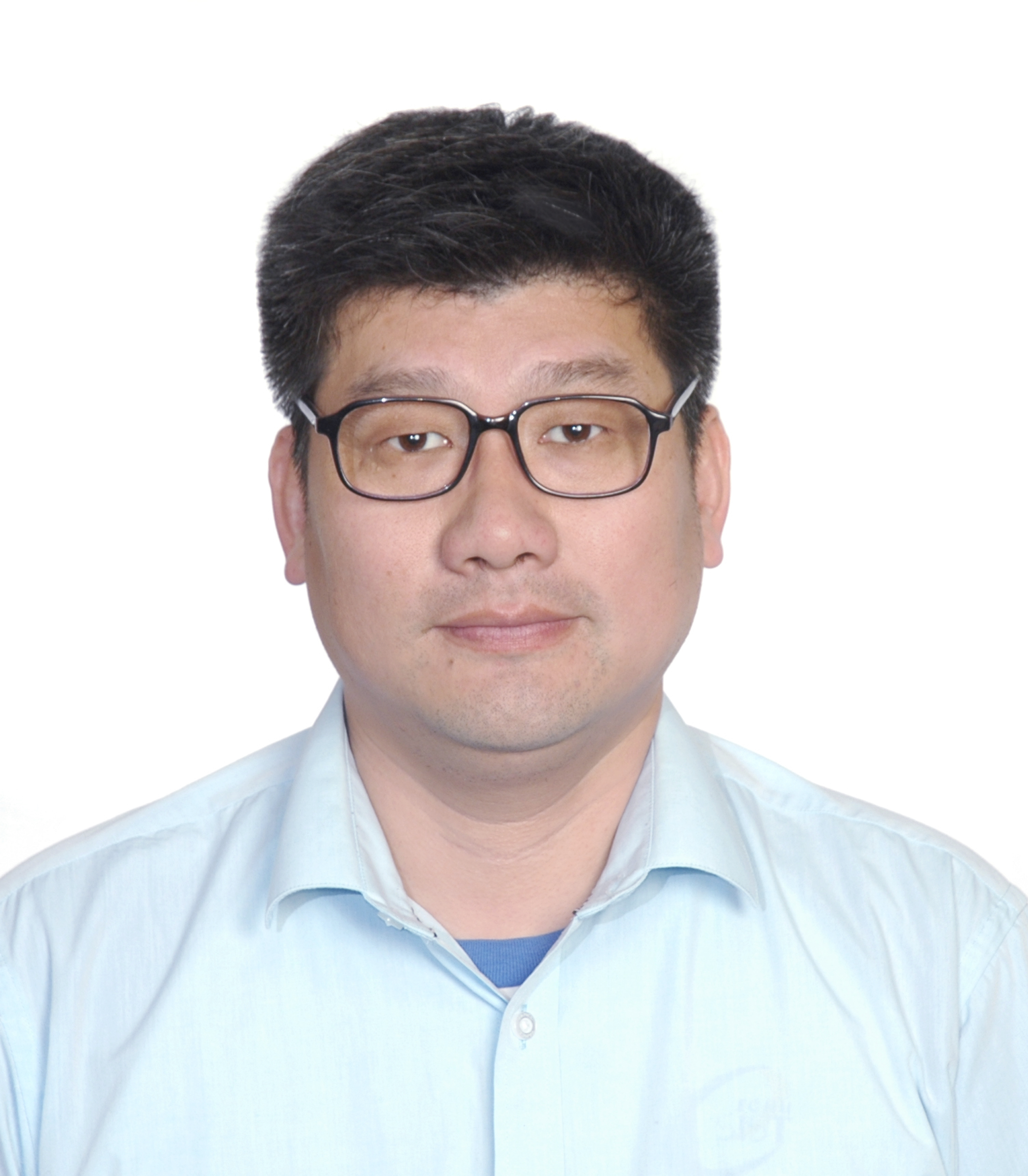}}]
{Haifeng Qian} is a professor at the Software Engineering Institute of East China Normal University, Shanghai, China. He received a BS degree and a master degree in algebraic geometry from the Department of Mathematics at East China Normal University, in 2000 and 2003, respectively, and the PhD degree from the Department of Computer Science and Engineering at Shanghai Jiao Tong University in 2006. His main research interests include network security, cryptography, and algebraic geometry. 
\end{IEEEbiography}

\end{document}